\documentclass[%
reprint,
superscriptaddress,
amsmath,
amssymb,
aps,
pre,
floatfix,
]{revtex4-2}

\usepackage{graphicx}% Include figure files
\usepackage{dcolumn}% Align table columns on decimal point
\usepackage{bm}% bold math
\usepackage{hyperref}% add hypertext capabilities
\usepackage[utf8]{inputenc}
\usepackage[T1]{fontenc}
\usepackage{mathptmx}
\usepackage{gensymb} 

\usepackage{amsmath}
\usepackage{amssymb}
\usepackage{graphicx}% Include figure files
\usepackage{dcolumn}% Align table columns on decimal point
\usepackage{bm}% bold math
\usepackage{multirow}
\usepackage{hyperref}
\usepackage{subfigure}
\usepackage{adjustbox}
\usepackage{tabularx}
\usepackage{url}
\DeclareSymbolFont{epsilon}{OML}{cmm}{m}{it}

\begin{document}

\preprint{APS/123-QED}

\title{Assessment of scoring functions for computational models of protein-protein interfaces}

\author{Jacob Sumner}
\thanks{These authors contributed equally to this work.}
\affiliation{Graduate Program in Computational Biology and Bioinformatics, Yale University, New Haven, Connecticut, 06520, USA}
\affiliation{Integrated Graduate Program in Physical and Engineering Biology, Yale University, New Haven, Connecticut, 06520, USA}
\author{Naomi Brandt}
\thanks{These authors contributed equally to this work.}
\affiliation{Integrated Graduate Program in Physical and Engineering Biology, Yale University, New Haven, Connecticut, 06520, USA}
\affiliation{Department of Physics, Yale University, New Haven, Connecticut, 06520, USA}
\author{Grace Meng}
\affiliation{Integrated Graduate Program in Physical and Engineering Biology, Yale University, New Haven, Connecticut, 06520, USA}
\affiliation{Department of Chemistry, Yale University, New Haven, Connecticut 06520, USA}
\author{Devon Finlay}
\affiliation{Integrated Graduate Program in Physical and Engineering Biology, Yale University, New Haven, Connecticut, 06520, USA}
\affiliation{Department of Physics, Yale University, New Haven, Connecticut, 06520, USA}
\author{Alex T. Grigas}
\affiliation{Graduate Program in Computational Biology and Bioinformatics, Yale University, New Haven, Connecticut, 06520, USA}
\affiliation{Integrated Graduate Program in Physical and Engineering Biology, Yale University, New Haven, Connecticut, 06520, USA}
\author{Andrés Córdoba}
\affiliation{Graduate Program in Computational Biology and Bioinformatics, Yale University, New Haven, Connecticut, 06520, USA}
\affiliation{Integrated Graduate Program in Physical and Engineering Biology, Yale University, New Haven, Connecticut, 06520, USA}
\author{Mark D. Shattuck}
\affiliation{Benjamin Levich Institute and Physics Department, The City College of New York, New York, New York 10031, USA}
\author{Corey S. O'Hern}
\affiliation{Graduate Program in Computational Biology and Bioinformatics, Yale University, New Haven, Connecticut, 06520, USA}
\affiliation{Integrated Graduate Program in Physical and Engineering Biology, Yale University, New Haven, Connecticut, 06520, USA}
\affiliation{Department of Mechanical Engineering, Yale University, New Haven, Connecticut, 06520, USA}
\affiliation{Department of Physics, Yale University, New Haven, Connecticut, 06520, USA}
\affiliation{Department of Applied Physics, Yale University, New Haven, Connecticut, 06520, USA}

\date{\today}

\begin{abstract}
An important goal of computational studies of protein-protein interfaces (PPIs) is to predict the binding site between two monomers that form a heterodimer. The simplest version of this problem is to rigidly re-dock the bound forms of the monomers, which involves generating computational models of the heterodimer and then scoring them to determine the most native-like models. PPI scoring functions have been assessed previously using rank- and classification-based metrics; however, these methods are sensitive to the number and quality of models in the scoring function training set. We assess the accuracy of seven PPI scoring functions by comparing their scores of computational models of PPIs to a measure of structural similarity to the x-ray crystal structure (i.e. the DockQ score) for a non-redundant set of heterodimers from the Protein Data Bank. For each heterodimer, we generate re-docked models uniformly sampled over DockQ and calculate the Spearman correlation between the PPI scores and DockQ. For some targets, the scores and DockQ are highly correlated; however, for many targets, there are weak correlations. Several physical features explain the difference between difficult- and easy-to-score targets. Strong correlations exist between the score and DockQ for targets with highly intertwined monomers and many interface contacts. We also develop a new score based on only two physical features that matches the performance of current PPI scoring functions. In addition, we address the more general problem of flexible-body docking by generating and docking intermediate monomer conformations between their bound and unbound forms. We score the docked models and find that the Spearman correlations between the PPI scores and DockQ decrease strongly as the monomers are deformed from their bound conformations. These results emphasize that PPI docking predictions can be improved by focusing on correlations between the PPI score and DockQ and incorporating more discriminating physical features into PPI scoring functions.
\end{abstract}

\maketitle

\section{\label{sec:intro}Introduction}

Proteins interact and form interfaces in numerous important biological processes, such as enzymatic catalysis of reactions, cytoskeletal organization, and immune recognition of pathogens. Identifying the binding interface between two proteins is crucial for understanding the function of protein complexes. The structure of protein-protein interfaces (PPIs) can be resolved via x-ray crystallography, nuclear magnetic resonance spectroscopy, and cryo-electron microscopy. However, experimentally resolving PPI structures is expensive and time-consuming. For example, there are more than $10^5$ PPIs in the human body, yet only $\sim 5\%$ of their structures have been experimentally resolved~\cite{orchard_mintact_2014, luck_reference_2020, drew_humap_2021, burke_towards_2023}. A key goal in the field of computational protein design is to predict whether two proteins form an interface and identify the location of the interface~\cite{dominguez_haddock_2003, torchala_swarmdock_2013, pierce_zdock_2014, yan_hdock_2017, marze_efficient_2018, lensink_challenge_2018, gainza_deciphering_2020, lensink_prediction_2021, lensink_impact_nodate, abramson_accurate_2024}. 

Determining the native binding interface in PPIs involves two steps: one must first know whether the two protein monomers interact and then identify their binding interface. In this work, we focus on the second step. In particular, we ask, given two monomers that are known to interact, can a near native structure be identified among a large pool of computational models of the heterodimer? The Critical Assessment of PRedicted Interactions (CAPRI) competition seeks to answer this question by regularly challenging research groups to predict the structures of novel experimentally determined PPIs from the amino acid sequence~\cite{lensink_prediction_2021, lensink_impact_nodate}. CAPRI, as well as many other computational studies of PPIs~\cite{dominguez_haddock_2003, moal_swarmdock_2010, pierce_zrank_2007, pierce_combination_2008, cheng_pydock_2007, huang_iterative_2008, park_simultaneous_2016, alford_rosetta_2017, olechnovic_voromqa_2017, zhou_goap_2011, pang_deeprank_2017, wang_protein_2020, wang_protein_2021, reau_deeprank-gnn_2023, xu_deeprank-gnn-esm_2024}, quantify the accuracy of models using the ``hit rate'' or ``success rate''. When using this metric, a scoring function is considered to be ``successful'' if any computational model in the top $N$ ranked models (where $N$ typically ranges from $5$ to $100$) is above a given arbitrary quality cutoff~\cite{mendez_assessment_2003, moal_scoring_2013, huang_exploring_2015, lensink_challenge_2018, lensink_prediction_2021, lensink_impact_nodate}. However, as discussed in detail below, the hit rate is highly sensitive to how many and which types of computational models are sampled. 

\begin{figure}[htbp]
    \includegraphics[width=\columnwidth]{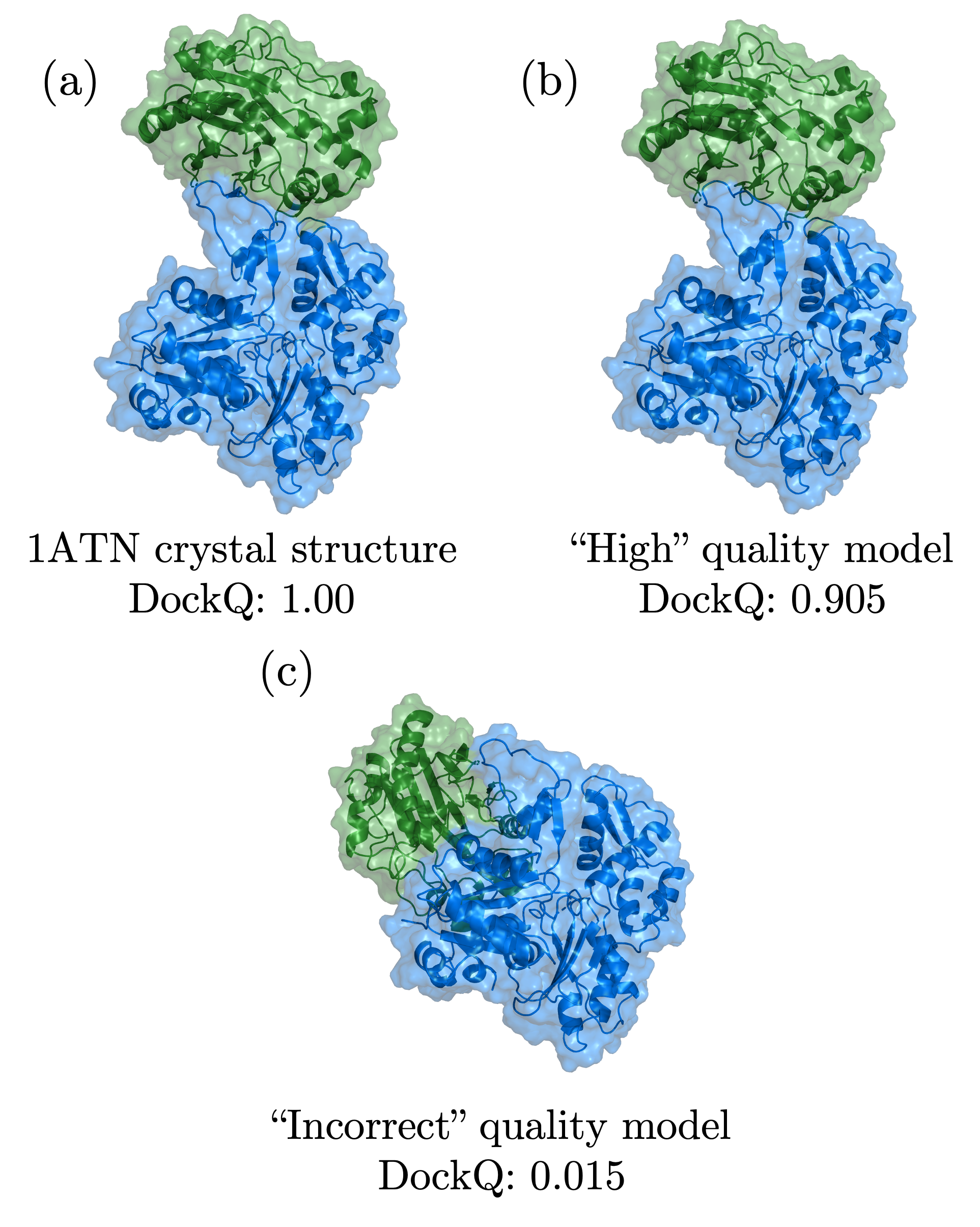}
    \caption{Visualization of (a) a heterodimer x-ray crystal structure (PDB ID: 1ATN), and (b) a ``high'' CAPRI quality rigid-body re-docked model and (c) an ``incorrect'' CAPRI quality rigid-body re-docked model for this target. The blue and green colors indicate the receptor and ligand of the heterodimer, respectively, and the DockQ scores are provided for the target and models.}
    \label{fig:protein_example}
\end{figure}

Rigid-body docking is the simplest form of the docking problem, wherein the monomers are rigid during docking. Thus, there are only six degrees of freedom to consider: three translations and three rotations of one monomer relative to the other. Rigid-body \textit{re-docking} considers the case where the monomers are in their native ``bound'' conformations. Thus, for PPIs with known bound conformations, sampling the full set of PPI computational models is straightforward, whereas the scoring of PPI computational models is a more difficult task. We can assess the quality of the top-ranked rigid-body re-docked models by comparing them to the x-ray crystal structure target using the continuous DockQ metric~\cite{basu_dockq_2016}, where $0 \leq {\rm DockQ} \leq 1$ and ${\rm DockQ} = 1$ indicates a computational model that is identical to the target. (See Fig. \ref{fig:protein_example} and Sec.~\ref{sec:MatMethods} for additional details.) The scoring of rigid-body re-docked structures is often considered to be a ``solved'' problem, with hit rates for {\it acceptable} (${\rm DockQ} \geq 0.23$) computational models exceeding 50\% in the top five scored models~\cite{xu_deeprank-gnn-esm_2024} using a common docking benchmark dataset with $\sim 200$ PPIs~\cite{vreven_updates_2015, guest_expanded_2021}. 

Despite the view that scoring rigid-body re-docked models is a settled problem, in this work we show that hit rate is sensitive to the number of computational models that are generated and how they are sampled. In addition, hit rate evaluates the scoring performance over a dataset of heterodimer targets, instead of evaluating the scoring performance for each target within the dataset. In contrast, the receiver operator characteristic (ROC) curve can determine the classification performance for a single target by measuring the area under the ROC curve (AUC). For the ROC curve, all scored models are assigned a label of ``true positive'' or ``true negative'' according to a threshold, or quality cutoff. However, similar to hit rate, AUC is also sensitive to how the computational models are sampled. 

% Figure 1

Determining the correlation between a predicted score and DockQ for each computational model is a more direct method for assessing the scoring accuracy of computational models. For example, when using a standard quality cutoff for a set of randomly sampled rigid-body re-docked computational models for a particular heterodimer target, we obtain an AUC of $0.959$, while the same data produces a Spearman correlation coefficient $|\rho| \approx 0.256$ \cite{mendez_assessment_2003}. The contradictory prediction accuracies reported for $\rho$ and AUC are caused by an imbalance of model quality in the data set, which is typical of current docking software. For random sampling of PPI models generated via rigid-body re-docking of high-resolution x-ray crystal structure targets, there are far fewer near-native models than low-quality models. Disparities in $\rho$ and AUC can also cause unreliable results when comparing the performance of existing PPI scoring functions. However, we have developed techniques to sample computational models uniformly over DockQ, which removes the disparities found in AUC and $\rho$ and enables reliable comparisons of PPI scoring performance among different targets and scoring functions. 

We consider a set of widely used, state-of-the-art PPI scoring functions to determine how well they distinguish near-native from low-quality rigid-body re-docked models of heterodimers. Rather than attempting a comprehensive benchmark of the $\gtrsim 100$ available scoring functions \cite{moal_scoring_2013}, we focus on representative methods used in recent CAPRI competitions: VoroMQA \cite{olechnovic_voromqa_2017}, ITScorePP \cite{huang_iterative_2008}, PyDock \cite{cheng_pydock_2007}, Rosetta \cite{park_simultaneous_2016, alford_rosetta_2017, marze_efficient_2018}, and ZRank2 \cite{pierce_zrank_2007, pierce_combination_2008,moal_scoring_2013, wang_protein_2020, wang_protein_2021, myung_csm-ab_2022}. These scoring functions span physics-based, statistically weighted, and hybrid methods. Because machine learning methods are frequently used in protein-structure prediction, we also assess two GNN-based scoring functions, Deeprank-GNN-ESM \cite{xu_deeprank-gnn-esm_2024} and GNN-DOVE \cite{wang_protein_2021}.

To evaluate the scoring functions, we generate rigid-body re-docked models that are uniformly sampled in DockQ for $84$ high-resolution heterodimer targets from the Protein Data Bank (PDB)~\cite{berman_protein_2000} and for a complementary set of $62$ non-redundant targets from the ZDOCK Benchmark-5.5 data set. For each target, we create 540{,}000 rigid-body re-docked computational models using ZDOCK3.0.2 \cite{chen_zdock_2003,pierce_accelerating_2011} and subsample them to obtain a set of models that uniformly samples the full range of DockQ. We then score each model with the seven scoring functions and evaluate their performance using the Spearman correlation coefficient $\rho$ with DockQ and the AUC across a range of DockQ cutoffs. When model sampling is uniform in DockQ, we find that $\mathrm{AUC} \approx -0.5\rho + 0.5$, and for roughly half of all targets $|\rho| < 0.70$, indicating that for most targets rigid-body re-docking is difficult for current scoring functions. By analyzing physical features of heterodimer interfaces, we have identified two key features that are well-correlated with DockQ: the number of interface contacts between the two monomers and the degree two which the two monomers are intertwined at the interface. We show that a simple support-vector regression (SVR) model with these features performs similarly to the most accurate of the seven other current scoring functions. 

As discussed earlier, rigid-body re-docking represents the simplest heterodimer docking problem. However, the more general task of PPI prediction involves flexible docking, i.e. first predicting the conformational changes in each monomer as they approach each other, and then identifying the interface between the two monomers in their bound conformations. In flexible docking, there is a large conformational search space that includes changes in the backbone and side chains between the unbound and bound conformations of the monomers. A conservative estimate of the number of degrees of freedom for a protein with $N$ amino acids is $4N$ with $2$ backbone and on average $\sim 2$ side chain dihedral angles per amino acid. Therefore, successful flexible docking depends on sufficiently sampling the large conformational space of the monomers and accurately scoring the docked conformations, even when the monomers are not in their bound conformations. Several research groups have created flexible docking algorithms~\cite{torchala_swarmdock_2013, dominguez_haddock_2003, gray_proteinprotein_2003, marze_efficient_2018} and participate in CAPRI. While the performance of flexible docking software has improved over the past $20$ years, it remains difficult to accurately predict the structure of heterodimers, which is a much simpler case than the large multimeric complexes that are frequently considered in CAPRI~\cite{lensink_impact_nodate}. Even deep-learning PPI structure prediction methods, such as AlphaFold-Multimer and AlphaFold 3~\cite{jumper_highly_2021, abramson_accurate_2024}, are unable to place $\sim 30\%$ of heterodimer PPIs within $10$\AA{} of their native structures~\cite{zhu_evaluation_2023}. In contrast, duplicate PPIs that have been crystallized by multiple groups possess root-mean-square deviations (RMSD) in the C$_{\alpha}$ atoms that are less than $1$\AA{} \cite{mei_analyses_2020}. 

We can take advantage of the reduction in the number of degrees of freedom provided by rigid-body re-docking by describing the flexible docking problem as a sequence of rigid-body docking calculations as the monomers deform incrementally from their bound to their unbound conformations. To assess the scoring function performance as a function of the RMSD of the monomers from their bound conformations, we generate monomer conformations that are linearly interpolated between the unbound and bound forms for a set of $15$ all-atom resolved heterodimers and their associated unbound monomers from the PDB. For each intermediate conformation, as well as the bound and unbound conformations,  we perform rigid-body docking, uniformly sample models over DockQ, and then score them. For each scoring function, we calculate $\rho$ between DockQ and the given score at each intermediate conformation. We find that $\rho$ (averaged over targets and averaged over scores) strongly decreases as the monomers are deformed from their bound conformations. 

The remainder of the article is organized as follows. In Sec.~\ref{sec:MatMethods}, we describe the methods used to assemble our dataset of models that are uniformly sampled over DockQ for the separate sets of $84$ and $62$ PPI targets. We also define the quantities that will be calculated to assess the PPI models including the hit rate, Spearman's rank correlation coefficient, interface RMSD (iRMSD), DockQ, and a new score based on physical features of protein interfaces. In Sec.~\ref{results}, we present our main findings. We first point out that the hit rate depends sensitively on the number and types of computational models that are used to calculate it and that measures of classification and correlation for assessing scoring function performance can differ when the computational models are not uniformly sampled in DockQ. We then evaluate the scoring function performance for rigid-body re-docked models under {\it uniform sampling} of DockQ for each PPI target. Finally, we generate monomer conformations that are linearly interpolated between their bound and unbound forms, and then dock, score them, and determine the Spearman correlation between DockQ and each score. We find that the Spearman correlation decrease strongly with the iRMSD of the monomers from their bound conformations. In Sec.~\ref{discussion}, we discuss the implications of these results for computational modeling of PPIs, and provide future research directions for integrating local physical features into graph-neural network scoring functions to improve their prediction accuracy. The article also includes four appendices. In Appendix~\ref{app:a}, we provide the scoring function performance for the ZDOCK Benchmark 5.5 dataset. Appendix~\ref{app:b} provides an overview of the protocols for evaluating each scoring function. In Appendix~\ref{app:c}, we examine the sensitivity of the hit rate on the number of sampled models. Finally, in  Appendix~\ref{app:d}, we visualize the DockQ and ZRank scoring landscapes for easy- and difficult-to-score PPI targets. We also give additional details for calculating the interface separability $S$ and number of interface contacts $N_c$ and compare the probability distributions $P(S)$ and $P(N_c)$ for the main dataset and the ZDOCK Benchmark 5.5 dataset of PPI targets. 

\section{Methods}
\label{sec:MatMethods}

In this section, we introduce the datasets, computational model-generation procedures, and analysis methods used to assess PPI scoring functions. We first curate multiple benchmark datasets of high-resolution x-ray crystal structures of heterodimer protein complexes. We then discuss how we generate large ensembles of rigid-body docked models using uniform sampling in DockQ, and describe the physics-based, statistical, and machine-learning scoring functions that we use to assess model quality. We quantify model accuracy using DockQ and evaluate prediction accuracy using hit-rate, interface RMSD, and the geometry of the DockQ landscape.

\subsection{Data Set Selection}
\label{subsec:data_set}

The main dataset is composed of 84 heterodimer x-ray crystal structures from the Protein Data Bank (PDB). (See Table \ref{tab:pdb_labels} in Appendix \ref{app:a} for a list of the structures.) The heterodimers were chosen according to the following criteria: all structures have resolution $\leq$ 3.5 \AA; no non-protein polymers, no metal ions, and no bound small molecules are present in the structure; both protein chains are at least $50$ amino acids long; the proteins are not artificially engineered and do not include amino acid mutations; and there are no missing amino acids in both chains. In addition, all heterodimers were filtered so that the structures have less than $20\% $ sequence similarity. Homodimers were excluded from the dataset to ensure that all targets occur as PPIs in a natural biological context and are not artifacts from crystallization \cite{schweke_discriminating_nodate}.

The $20\%$ sequence similarity cutoff did not include both monomer chains. In the $84$ target dataset, there are four targets that contain ubiquitin as a monomer: 1WRD, 2HTH, 3K9P, and 7JXV. There are two targets that both contain the same ubiquitin conjugating enzyme: 1J7D and 4DHI. There are two targets that both contain the same SUMO conjugating enzyme: 2GRN and 2PE6. Finally there are two targets that both contain the same PAPD chaperone protein: 1PDK and 2WMP. While a few of the monomers are similar, the interfaces in each of the $84$ heterodimers are unique. 

We also considered heterodimer targets from the ZDOCK Benchmark-5.5 dataset \cite{vreven_updates_2015} as a separate dataset to corroborate our results. We filtered the ZDOCK Benchmark-5.5 dataset to obtain $67$ structures that were true heterodimers. We then removed the structures that overlapped with the original $84$ target dataset, which yielded $62$ additional structures. These targets are listed in Table \ref{tab:bm55_labels} in Appendix \ref{app:a}.

In addition, we constructed a smaller dataset of $15$ x-ray crystal structure heterodimers that also possess x-ray crystal structures for both monomers in their unbound conformations to study rigid-body docked models as a function of the RMSD from the bound conformations. $13$ of the $15$ structures (PDB IDs: 1FLE, 2A5T, 2OZA, 2HRK, 2Z0E, 1ACB, 1ZLI, 2C0L, 1BKD, 2O3B, 1ATN, 1FQ1, and 1R8S) are included in the ZDOCK Benchmark 5.5 dataset~\cite{vreven_updates_2015}.  For the remaining two structures (PDB IDs: 1UGH and 3K9P), we selected monomers with resolution $\leq3.5\mathring{\text{A}}$ with 100\% sequence overlap with the monomers found in the x-ray crystal structure of the heterodimer.  The complexes and individual monomers for this dataset are given in Table \ref{tab:irmsd_pdbids} in Appendix \ref{app:a}.

\subsection{Model Generation and Sampling}
\label{subsec:sampling}

The computational models were generated for each of the heterodimer targets using ZDOCK 3.0.2 with 6-degree Euler angle increments, which yields 54,000 models per ZDOCK run. ZDOCK was run $10$ times per target. For each run, the monomers were randomly rotated and translated to ensure that all computational models were distinct. We calculate $0 \le {\rm DockQ} \le 1$ for all 540,000 models for each target and sort them into $20$ evenly spaced DockQ bins to achieve uniform sampling in DockQ. For each target, a maximum of $50$ models was selected from each DockQ bin, yielding a maximum of $1000$ models per target. In addition, to provide even sampling of ``negative'' models, $1000$ models with ${\rm DockQ} < 0.23$ were randomly selected from the models that were not already selected. All selected models had hydrogens added to them using Reduce \cite{davis_molprobity_2007} from the PHENIX software package \cite{adams_phenix_2011} and were then relaxed using Rosetta v3.12 \cite{park_simultaneous_2016, alford_rosetta_2017}. After relaxation, the set of models was balanced so that each heterodimer target has an equal number of positive (${\rm DockQ} \ge 0.23$) and negative (${\rm DockQ} < 0.23$) models. 

To generate intermediate monomer conformations between the bound and unbound forms, we required both structures to have the same number of heavy atoms. We found that there were small differences in the number of resolved amino acids in the bound and unbound monomer pairs, and thus we performed sequence alignment between the pairs. After alignment, trailing amino acids at the N- or C-termini of the structures were removed to ensure the same protein length. If monomer pairs exhibited single nucleotide polymorphisms, the differing amino acids were not included in the analyses. We restored missing or partial residues using OpenMM's PDBFixer module \cite{eastman_openmm_2024}.

We calculate the displacement matrix ${\overline D}$ between the bound and unbound conformations of a given monomer with $N_h$ heavy atoms as
\begin{equation}
{\overline D}=
\begin{pmatrix}
\vec{r}_{U,1}-\vec{r}_{B,1}\\
\vdots \\
\vec{r}_{U,N_h}-\vec{r}_{B,N_h} 
\end{pmatrix},
\label{eq:interp_stepsize}
\end{equation}
where $\vec{r}_{B,i}$ and $\vec{r}_{U,i}$ are the positions of heavy atom $i$ in the bound and unbound conformations of the monomer, respectively. We obtain intermediate monomer conformation $I_n$ by linearly interpolating between the bound and unbound conformations: 
\begin{equation}
I_n=
\begin{pmatrix}
\vec{r}_{B,1 }\\
\vdots \\
 \vec{r}_{B,N_h } 
\end{pmatrix}
+\frac{n}{N_I}{\overline D},
\label{eq:interp_image}
\end{equation}
where $0\le n < N_I$ and $N_I=6$ is the number of images used in this study, such that the interval size $|{\overline D}|/N_I \lesssim1\mathring{\text{A}}$ between images. All images were relaxed using Rosetta v3.12 \cite{park_simultaneous_2016, alford_rosetta_2017}. For each image, we rigid-body dock the pairs of monomers at the same $n$ using the docking protocol described above.

\subsection{Assessing Protein Models}
\label{subsec:model_scoring}

The $\sim 1000$ uniformly sampled models for each target were scored using seven representative scoring functions in recent CAPRI competitions: ZRANK2~\cite{pierce_zrank_2007,pierce_combination_2008}, ITScorePP~\cite{huang_iterative_2008}, Rosetta 3.12 (using the {\tt ref2015} weight function)~\cite{alford_rosetta_2017, marze_efficient_2018}, VoroMQA version 1.26 \cite{olechnovic_voromqa_2017}, PyDock3~\cite{cheng_pydock_2007}, Deeprank-GNN-ESM~\cite{xu_deeprank-gnn-esm_2024}, and GNN-DOVE~\cite{wang_protein_2021}. See Appendix \ref{app:b} for more information on how each scoring function was implemented.

\subsubsection*{Hit Rate Plots}
\label{subsec:hit_rate}

The scoring accuracy can be quantified using the hit rate fraction $h^R$, which is defined as 
\begin{equation}
\label{eq:hit_rate}
    h^R = \frac{1}{{\cal N}} \sum_{k=1}^{\cal N} \frac{1}{N_{m}^{k,+}} \sum_{i=1}^{R} \Theta({{\rm DockQ}_{k}^{i}} - {\rm DockQ}_0),
\end{equation}
where ${\cal N}$ is the number of targets in the dataset, $R$ is the maximum rank considered (usually $1 \leq R \leq 100$) in the set of models ordered by a given PPI score, $N_{m}^{k,+}$ is the number of positives in the set of models up to rank $R$, ${\rm DockQ}_{k}^{i}$ is the DockQ value of the $i$th model for target $k$, ${\rm DockQ}_0=0.23$ is the DockQ cutoff that defines a ``positive'' model, and  $\Theta(\cdot)$ is the Heaviside step function which is nonzero for ${\rm DockQ}_k^i > {\rm DockQ}_0$.

The effective hit rate plot can also be calculated for each target $k$, individually:
\begin{equation}
    h_k^R = \frac{1}{N_s} \sum_{l=1}^{N_s} \frac{1}{N_{m}^{k,l,+}} \sum_{i=1}^{R} \Theta({{\rm DockQ}_{k,l}^{i}} - {\rm DockQ}_0).
\label{eq:eff_hitrate}
\end{equation}
For each target, multiple samples with sizes $N_m$ are drawn from the total number of models $N_t$ ranging from $N_m=500$ to $15$,$000$. $N_s=1000$ samples are drawn for each value of $N_m$, and $h_k^R$ is calculated at each rank $i$ using Eq.~\ref{eq:eff_hitrate}, where ${\rm DockQ}_{k,l}^{i}$ is the DockQ value for the $l$th sample at rank $i$ for the $k$th target. $N_{m}^{k,l,+}$ represents the set of ``positive'' models in the random sample of $N_m$ models. We only consider up to rank $R=100$, and the models are scored and ordered using the Rosetta scoring function. 

\subsubsection*{Spearman Correlation Coefficient}

The Spearman's (rank) correlation coefficient $\rho$ is a parameter-free measure of monotonic association between two variables~\cite{spearman_1904}. Rather than comparing the raw value of the two variables, the Spearman correlation is calculated by replacing each variable with its rank ordering and then evaluating the Pearson correlation coefficient between those two rank ordered sets. Thus, the Spearman correlation coefficient can compute correlations between any monotonic change in two variables that are linearly or non-linearly associated, regardless of the scale of the original variables. The Spearman correlation satisfies $-1 \le \rho \le 1$, where $|\rho| = 1$ corresponds to a monotonic relationship between the variables and $\rho = 0$ indicates that there is not a monotonic association.

Given paired observations $({\rm DockQ}_i,s_i)$ for $i=1,\ldots,N_m$ computational models for a given PPI target, let $R({\rm DockQ}_i)$ and $R({s}_i)$ denote the ranks of ${\rm DockQ}_i$ and ${s}_i$ among the $N_m$ models separately.  ${s}_i$ for model $i$ indicates one of the scoring functions discussed in Sec.~\ref{subsec:model_scoring}. The Spearman correlation coefficient is
\begin{equation}
\begin{split}
\rho = 
\frac{
\sum_{i=1}^{N_m}
\left(R({\rm D}_i) - \overline{R({\rm D}_i)}\right)
\left(R({s}_i) - \overline{R({s_i})}\right)
}{
\sqrt{
\sum_{i=1}^{N_m}
\left(R({\rm D}_i) - \overline{R({\rm D}_i)}\right)^2
}
\sqrt{
\sum_{i=1}^{N_m}
\left(R({s}_i) - \overline{R({s_i})}\right)^2
}
},
\end{split}
\end{equation}
where ${\rm D}_i \equiv {\rm DockQ}_i$, and $\overline{R({\rm D}_i)}$ and $\overline{R({s}_i)}$ are the mean ranks.

\subsubsection*{Interface Root-Mean-Square Deviation (iRMSD)}
\label{subsec: irmsd}

To calculate iRMSD, we identify the interface residues using the relative Solvent Accessible Surface Area (rSASA)  \cite{grigas_using_2020}. The interface residues are defined as any residues with $\Delta\rm{rSASA}<0$, where $\Delta\rm{rSASA}=rSASA_B-rSASA_M$, $\rm{rSASA}_B$ denotes the rSASA of each residue in the bound state, and $\rm{rSASA}_M$ represents the rSASA of each residue of the isolated monomers in their bound conformations. Thus, any residue that exhibits an increase in rSASA between the dimer and monomer forms must be in contact with a residue in the opposing monomer and is considered to be at the interface. This definition of interface residues is more accurate than that in the ZDOCK Benchmark-5.5 dataset, which defines interface residues as those with at least one heavy atom within 10\AA\ of any heavy atoms in the opposing monomer \cite{vreven_updates_2015}. 

To calculate the interface RMSD, we first align the interface residues of each monomer to their respective bound conformations. We then define the mean-squared displacement of interface residues on each monomer $\beta=1$,$2$ relative to their bound forms: 
\begin{equation}
d_{B,\beta}=\sum_{i=1}^{N_{\beta}}(
\vec{r}_{B,\beta,i}-\vec{r}_{\beta,i})^2,
\label{eq:sq_dist1}
\end{equation}
where $N_{\beta}$ is the number interface residues in monomer $\beta$, $\vec{r}_{B,\beta,i}$ and $\vec{r}_{\beta,i}$ are the positions of C$_\alpha$ atom $i$ in monomer $\beta$ at the interface in the bound structure $S_B$ and any other monomer conformation, respectively. We define the iRMSD as 
\begin{equation}
{\rm iRMSD}=\sqrt{\frac{d_{B,1}+d_{B,2}}{N_{1}+N_{2}}}.
\label{eq:irmsd}
\end{equation}

\subsubsection*{DockQ Metric for Structural Similarity}
\label{subsec:DockQ}

We employed DockQ as the ground truth score for PPI structures~\cite{basu_dockq_2016}. DockQ is a measure of the structural similarity between each model and the x-ray crystal structure target, where $0 \leq {\rm DockQ} \leq 1$. ${\rm DockQ} = 1$ indicates that a model is identical to the target x-ray crystal structure, and ${\rm DockQ} = 0$ indicates that a model is very dissimilar to the target x-ray crystal structure. DockQ is the average of three terms: ${\rm DockQ} = (F_{nat} + 1/(1+({\rm LRMSD}/ d_1)^2 + 1/(1+({\rm iRMSD}/ d_2)^2)/3$. $F_{nat}$ is the fraction of ``native contacts'' $N_c^{\rm DockQ}$ from the x-ray crystal structure that also appear in the computational model. Native contacts for DockQ are defined between pairs of heavy atoms on different monomers that are separated by less than 5 \AA. The ligand RMSD (LRMSD) is the RMSD of the ligand backbone atoms when the model and the x-ray crystal structure are aligned on the ``receptor'' monomer. For DockQ, heavy atoms are considered to be at the interface if they are less than 10 \AA~from a heavy atom on an opposing monomer. The factors $d_1=8.5$ \AA~and $d_2=1.5$ \AA~scale the LRMSD and iRMSD so that $0 \le {\rm DockQ} \le 1$.

\subsection{Assessing physical properties of protein interfaces}
\label{assess}

\subsubsection*{DockQ Landscape for Heterodimers}
\label{subsec:Sphere plot}

The DockQ landscape for PPI models contains a series of points (in spherical coordinates) that represent the centers of mass (COM) of the ligand C$_{\alpha}$ atoms of computational models for a given target. We assign the receptor and ligand using the same definition as that in the target PDB file. All models are aligned so that the receptor in each model has minimal C$_{\alpha}$-RMSD to the receptor in the x-ray crystal structure of the target. Each model is translated so that the center of mass of the receptor is at the origin. In the DockQ landscape, the ligand COM position for each model is placed at a distance $R$ between the centers of mass of the receptor and ligand in the x-ray crystal structure of the target. 

\subsubsection*{Relative Anisotropy of the DockQ Landscape}
\label{subsec:relative_aniso}

The anisotropy $\kappa^2$ of the DockQ landscape is calculated using the eigenvalues of the moment of inertia matrix formed from the centers of mass of the ligands of the computational models. The moment of inertia matrix for a single target is defined as 
\begin{equation}
\label{eq:inertia}
    I_{\alpha \beta} = \frac{1}{N_m}\sum_{i=1}^{N_m} {\rm DockQ}^3_i \left( r_i^2 \delta_{\alpha \beta} - r^i_{\alpha }r^i_{\beta}\right),
\end{equation}
where $N_m$ is the number of models, $\alpha$,$\beta=x$,$y$,$z$, $\delta_{\alpha \beta}$ is the Kronecker delta, $r^i_{\alpha}$ is $\alpha$-component of the position of the $i$th point, and $r_i^2 = x_i^2 + y_i^2 + z_i^2$. The moment of inertia is weighted by $\rm{DockQ}^3_i$ to emphasize native-like models. The components of the moment of inertia matrix $I$ are
\begin{equation}
I_{\alpha \beta} =
\begin{pmatrix}
I_{xx} & I_{xy} & I_{xz} \\
I_{yx} & I_{yy} & I_{yz} \\
I_{zx} & I_{zy} & I_{zz}
\end{pmatrix}.
\label{eq:intertia_tensor}
\end{equation}
$I$ can be expressed as
\begin{equation}
I = U^{t} \Lambda U,
\label {eq:eigen}
\end{equation}
where $U$ represents an orthogonal matrix whose columns are the normalized eigenvectors of $I$ and $\Lambda$ is the diagonal matrix of eigenvalues, 
\begin{equation}
\Lambda =
\begin{pmatrix}
\lambda_1 & 0 & 0 \\
0 & \lambda_2 & 0 \\
0 & 0 & \lambda_3
\end{pmatrix},
\label{eq:eigenvalue_matrix}
\end{equation}where $\lambda_1 \leq \lambda_2 \leq \lambda_3$. The eigenvalues of $I$ can be used to calculate the relative anisotropy: 
\begin{equation}
\label{eq:relative_anisotropy}
    \kappa^2 = 1 - 3 \frac{\lambda_1 \lambda_2 + \lambda_2 \lambda_3 + \lambda_1 \lambda_3}{(\lambda_1 + \lambda_2 + \lambda_3)^2}.
\end{equation}

\subsubsection*{Physical Features of Proteins}
\label{subsec:physical_features}

We calculate the separability $S$ of each interface using a degree-3 polynomial support vector machine (SVM) from the SciKit-Learn package in python~\cite{pedregosa_scikit-learn_2011}. The separability satisfies $0.5 < S < 1.0$, where the SVM surface can perfectly separate the ligand and receiptor for $S=1.0$, while $S=0.5$ corresponds to a highly intertwined interface. 

The number of interface contacts $N_c$ is defined as the number of heavy atom pairs in different protein chains with separations less than $d_c=4.5$ \AA. We also calculated the number of interface contacts for all possible distance cutoffs from $d_c=1$-$30$ \AA~with an increment of $0.1$ \AA. We find that $d_c=4.5$ \AA~ yielded the largest Spearman correlation between $N_c$ and DockQ for the models that are uniformly sampled in DockQ. 

\section{Results}
\label{results}

Our goal is to assess the accuracy of current scoring functions for computational models of rigid-body re-docked protein–protein interactions generated from the monomers of high-resolution x-ray crystal structure heterodimers. In Sec.~\ref{subsec:results_hitrate}, we first show that hit rate plots are an unreliable metric for evaluating scoring function performance when PPI scores are weakly correlated with DockQ or when model quality is insufficiently sampled. We address the latter issue in Sec.~\ref{subsec:scoring_performance} by uniformly sampling models over the full range of DockQ for each target and using both classification and correlation based metrics for assessing scoring accuracy. Using this sampling and scoring assessment methodology, we evaluate six representative PPI scoring functions from recent CAPRI competitions, ZRank2, ITScorePP, Rosetta, PyDock, VoroMQA, GNN-DOVE, and Deeprank-GNN-ESM~\cite{lensink_impact_nodate}, plus an additional SVR model that we constructed based on key physical features of protein interfaces. We find that the scoring performance varies strongly across targets, and we therefore investigate the physical properties that determine scoring performance. In Sec.~\ref{subsec:results_physical_features}, we calculate the number of interfacial contacts $N_c$ and the degree of interface separability $S$ for all PPI targets, and show that these features are strongly correlated with the scoring difficulty. We combine these features into an SVR model and show that it matches or exceeds the scoring performance of current state-of-the-art scoring functions in Sec.~\ref{subsec:results_svr}. Finally, in Sec.~\ref{subsec:flexible}, we examine the scoring performance for flexibly docked structures by docking monomer conformations that have been interpolated between their respective bound and unbound forms and show that the scoring function performance decreases rapidly as the iRMSD increases.

\subsection{The hit rate metric can be an unreliable measure of scoring function efficacy}
\label{subsec:results_hitrate}
% Figure 2
\begin{figure*}[htbp]
    \centering
    \includegraphics[width=0.95\textwidth]{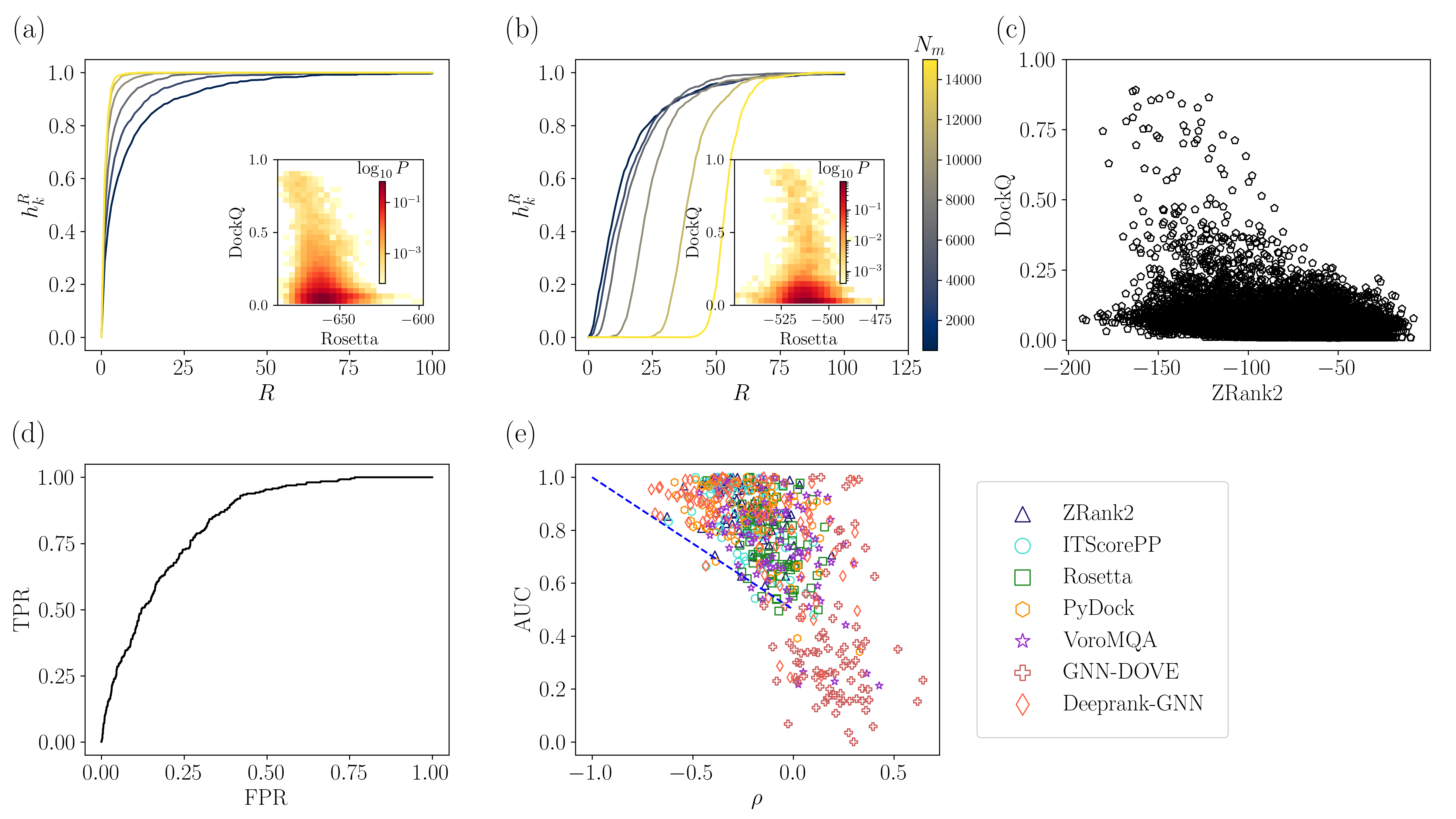}
    \caption{The effective hit rate fraction $h_k^R$ is plotted versus the maximum considered rank $R$ of the models ordered according to their Rosetta score for increasing numbers of computational models $N_m$ (from blue to yellow) for (a) 3RCZ and (b) 3YGS. The insets show the probability distribution $P({\rm DockQ},{\rm Rosetta})$ for the computational models to have given DockQ and Rosetta scores when exhaustively sampling the computational models for heterodimer targets 3RCZ and 3YGS. The color scale from light yellow to dark red indicates increasing probability. (c) Scatterplot of ZRank2 versus DockQ for PDB 3RCZ with Spearman correlation $\rho = -0.303$. (d) ROC curve for the data in (c) with true positive cutoff DockQ $\geq 0.23$ and ${\rm AUC} = 0.825$. (e) The Spearman correlation $\rho$ between DockQ and each of the seven scoring functions plotted versus the AUC for all targets. Each color represents a different target and the blue dotted line represents ${\rm AUC}=-0.5\rho+0.5$.}
    \label{fig:sampling_hitrate}
\end{figure*}

Currently, the most commonly used method to assess PPI scoring function accuracy is the hit rate plot~\cite{olechnovic_voromqa_2017, wang_protein_2020, wang_protein_2021, pang_deeprank_2017, reau_deeprank-gnn_2023}, which measures the fraction of targets with at least one ``positive'' (i.e. models with DockQ $\geq 0.23$) in the top $i$ ranked models (out of a total of $N_t$ models) scored by a given scoring function averaged over all targets. (Models with DockQ $<0.23$ are referred to as ``negatives.'') However, the hit rate plot is extremely sensitive to $N_t$ when there are weak correlations between the ground truth score (i.e. DockQ) and scores from PPI scoring functions. To illustrate the sensitivity of the hit rate metric on the number of models that are scored, we calculate the {\it effective} hit rate fraction $h_k^R$ for each target. (See Eqs. \ref{eq:hit_rate} and \ref{eq:eff_hitrate} for more information.) We first generate $N_{t}=108$,$000$ computational models using ZDOCK for each target. We then randomly draw $N_m$ models from this collection $N_{s}$ times. The $N_m$ models are ranked using the Rosetta scoring function from lowest to highest, $i=1$,$\ldots$,$N_m$. The effective hit rate fraction at $i$ is the fraction of the $N_s$ draws that have at least one positive model for ranks less than or equal to $i$. In Fig. \ref{fig:sampling_hitrate} (a) and (b), we show the dependence of the effective hit rate on the sampling size. We generate $N_{t}$ models generated by ZDOCK for two example targets, 3RCZ~\cite{prudden_dna_2011} and 3YGS~\cite{qin_structural_1999}, respectively. The log of the probability distribution $P({\rm DockQ},{\rm Rosetta})$ is shown in the inset of each panel, which highlights the likelihood of a model to have a given DockQ and Rosetta score. The magnitude of the Spearman correlation coefficient $|\rho|$ for 3YGS between DockQ and the Rosetta score ($|\rho| = 0.35$) is much smaller than that for 3RCZ ($|\rho| = 0.62$). (Details on how the Spearman correlation coefficient is calcuated is included in Section \ref{subsec:model_scoring}.) For target 3RCZ, we show in Fig.~\ref{fig:sampling_hitrate} (a) that $h_k^R$ improves as $N_m$ increases, which suggests that a larger sample size can lead to an increase in $h_k^R$ for some targets. However, in Fig. \ref{fig:sampling_hitrate} (b), $h_k^R$ decreases as $N_m$ increases for 3YGS. In particular, as $N_m$ approaches $N_t$, $h_k \rightarrow 1$ for 3RCZ, while for 3YGS, $h_k \rightarrow 0$. These results emphasize that the effective hit rate fraction is sensitive to the number of models $N_m$. Similar results for the hit rate fraction are found for all other scoring functions that we tested. 

Why does $h_k^R$ tend to the limiting values of $0$ or $1$ as $N_m \rightarrow \infty$?  An important difference between the targets 3RCZ and 3YGS is that Rosetta assigns lower scores to negative models for 3YGS than those for positive models. Thus, it is important to quantify the fraction of computational models $\mu$ that are negative and have a PPI score (e.g. Rosetta score) that is lower than the minimum value for all positive models. To understand the dependence of the effective hit rate fraction on $\mu$, we generated a large synthetic dataset of ``docked models'' for a PPI target with a Spearman correlation $|\rho| \approx 0.1$ between ``DockQ'' and ``PPI score''. We balance the synthetic dataset so that it has an equal number of positive and negative models, assuming again that the positive models have ``DockQ'' $\geq 0.23$. In Fig. \ref{fig:supp_synthetic_hr} in Appendix \ref{app:c}, we show that the effective hit rate fraction tends toward zero for all $\mu > 0$, which emphasizes the unreliability of the hit rate fraction when considering the large number of samples limit.

% Figure 3
\begin{figure*}[htbp]
\centering
\includegraphics[width=\textwidth]{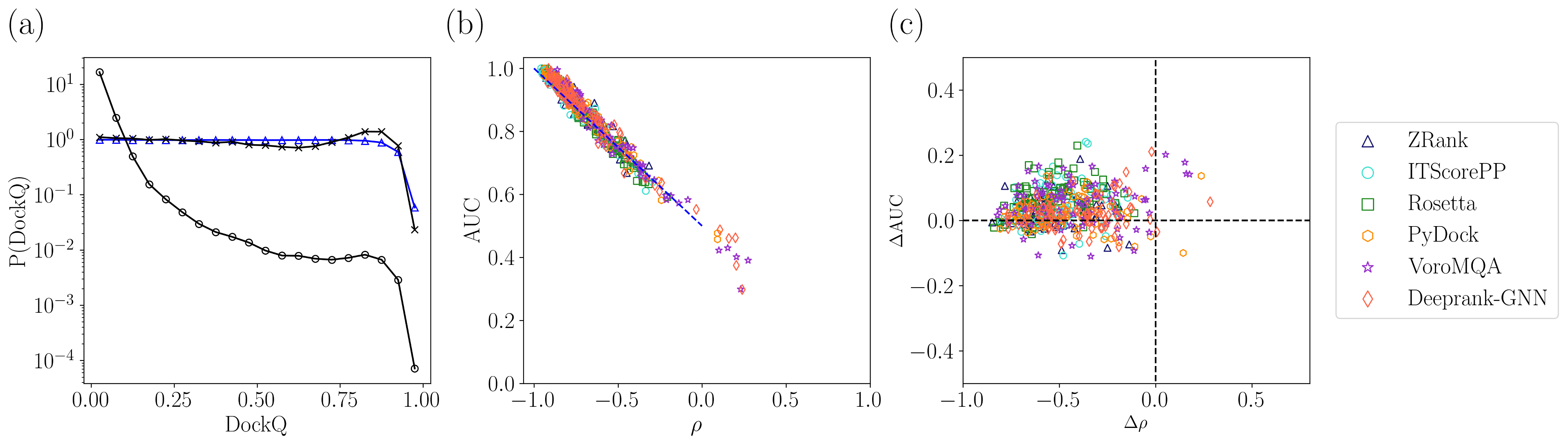}
\caption{Improvements in the evaluation of PPI scoring functions due to sampling models uniformly in DockQ. (a) Probability distribution $P({\rm DockQ})$ of DockQ for computational models for all targets obtained by exhaustive sampling of models (gray circles), uniformly sampling DockQ before energy minimization (blue triangles), and uniformly sampling DockQ after energy minimization (black crosses). (b) ${\rm AUC}$ plotted versus $\rho$ for models that uniformly sample DockQ (after energy minimization) for each scoring function. The blue dotted line gives ${\rm AUC}=-0.5\rho+0.5$. (c) $\Delta \rho = \rho_{u} - \rho_{e}$ plotted versus $\Delta {\rm AUC} = {\rm AUC}_u - {\rm AUC}_e$, where $\rho_{u}$ ($\rho_e$) is the Spearman correlation for the models that were uniformly (exhaustively) sampled in DockQ and ${\rm AUC}_u$ (${\rm AUC}_e)$ is the area under the ROC curve for the models that were uniformly (exhaustively) sampled.} 
\label{fig:compare_supersampled}
\end{figure*}

Another metric that can be used to assess PPI scoring function performance is the area under the ROC curve (AUC). $0.5 < {\rm AUC} < 1$ indicates how well a scoring function classifies models as positive versus negative. AUC values of $0.5$ and $1$ indicate random and perfect classification, respectively. In Fig.~\ref{fig:sampling_hitrate} (c), we show a scatterplot of DockQ versus ZRank2 scores for {\it all} models of PDB 3U82 (i.e. not uniformly sampled in DockQ) with Spearman correlation $\rho = -0.191$. The ROC curve for the same target (with ${\rm AUC} = 0.707$) is shown in Fig. \ref{fig:sampling_hitrate} (d). We compare $\rho$ and AUC for the $84$ targets in the main dataset in Fig. \ref{fig:sampling_hitrate} (e). These results show inconsistency between the classification and correlation metrics. For many targets, the ${\rm AUC} \sim 1$, while $|\rho| \leq 0.5$. For example, GNN-DOVE inconsistently scores high DockQ models, either assigning them the highest or lowest scores. Thus, we omit GNN-DOVE from further analyses. For PPI scoring functions, why is the magnitude of the Spearman correlation coefficient so low, yet the AUC indicates accurate classification? 

\subsection{Scoring performance is reliable after uniformly sampling over DockQ}
\label{subsec:scoring_performance}
The disparity between $\rho$ and AUC likely stems from the overabundance of poor-quality models and lack of high-quality models typically generated by PPI docking methods. To address both issues, we generate a dataset of $\sim 50$ models (sampled from a larger pool of $540$,$000$ models per target) for each DockQ bin with width $0.05$ (Fig.~\ref{fig:compare_supersampled} (a)). After energy minimizing the models with Rosetta, the distribution of model quality is nearly uniform except for the highest few bins of DockQ. We then re-balance the dataset so that half of the models are positive and half are negative by adding negative models to the energy minimized dataset. This uniform sampling procedure gives $\sim 1400$ models per target. We then calculate $\rho$ and the AUC for the uniformly sampled dataset and find that the ${\rm AUC}$ obeys ${\rm AUC} \approx -0.5 \rho + 0.5$ as shown in Fig. \ref{fig:compare_supersampled} (b). Since $\rho$ and ${\rm AUC}$ are linearly related for the uniformly sampled data, we focus on $\rho$ for the remaining analyses. 

% Figure 4
\begin{figure*}[htbp]
\centering
\includegraphics[width=0.95\textwidth]{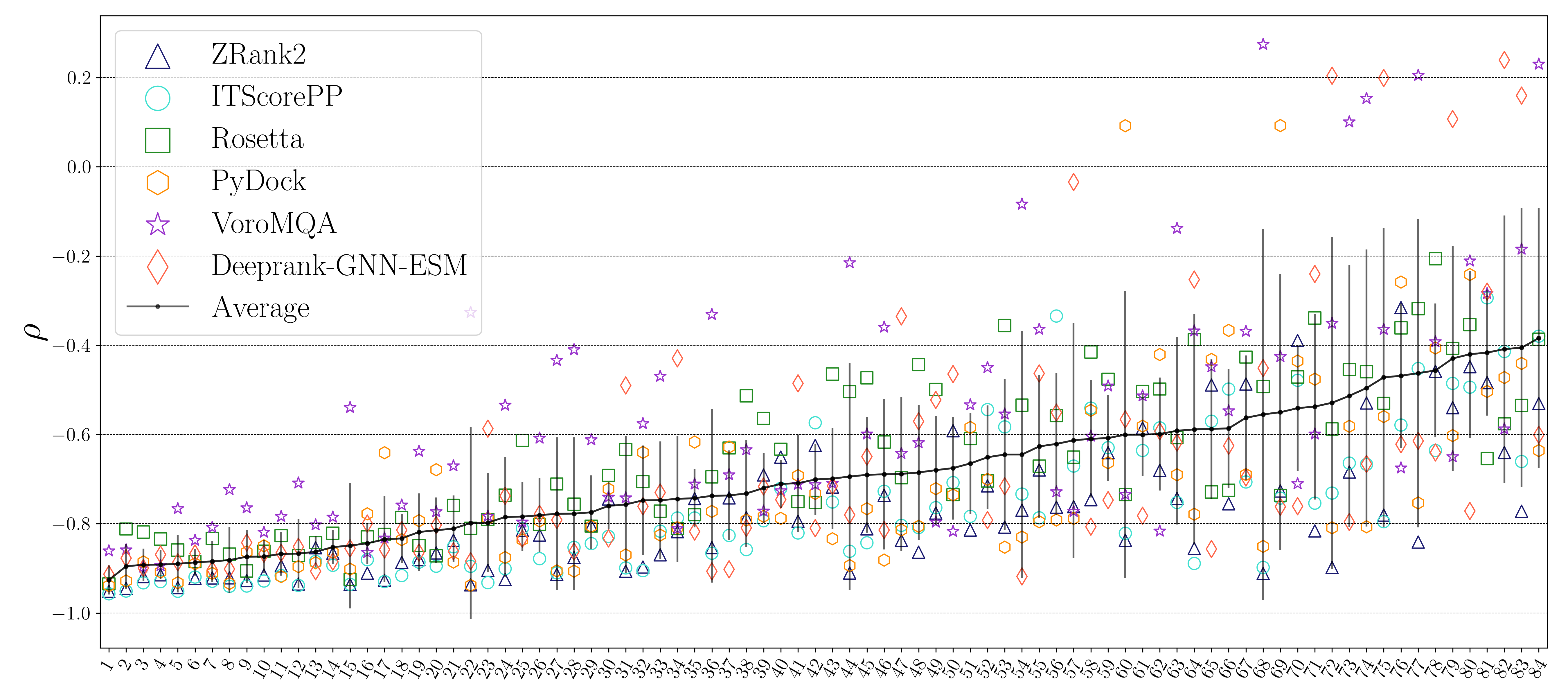}
\caption{The Spearman correlation $\rho$ between DockQ and PPI score for each target and scoring function. We order the targets by increasing $\langle \rho \rangle$ (black line), averaged over the six scoring functions listed, from left to right. (Note that since $\rho < 0$ increasing $\rho$ implies decreasing the magnitude of the correlations.) The standard deviations of the scores for each target are indicated by the black vertical lines. The numbers on the horizontal axis correspond to PPI targets listed in Table \ref{tab:pdb_labels} in Appendix \ref{app:a}.}
\label{fig:per_target_spearman_scores}
\end{figure*}

To quantify the change in the evaluation metrics of the scoring functions after uniformly sampling models in DockQ, we compared $\rho$ and AUC obtained from exhaustive versus uniform sampling. We define $\Delta \rho = \rho_{\rm{u}} - \rho_{\rm{e}}$ and $\Delta\rm{AUC} = \Delta\rm{AUC}_{\rm{u}} - \rm{AUC}_{\rm{e}}$, where  ``u'' and ``e'' represent the uniform and exhaustively sampled datasets. In Fig. \ref{fig:compare_supersampled} (c), we show a scatterplot of $\Delta \rho$ versus $\Delta \rm{AUC}$. We find that most of the data points are found in the second quadrant, which indicates improvements in both correlation and classification metrics following uniform sampling of models in DockQ. (Increased correlations correspond to $\Delta \rho <0$ and improved classification corresponds to $\Delta \rm{AUC} >0$.) Some targets have a negative $\Delta \rm{AUC}$ or a positive $\Delta \rho$, which likely stems from under-sampling of positive models for targets in the exhaustively sampled dataset. These results emphasize that after uniformly sampling models in DockQ, $\rho$ is a consistent metric for evaluating the performance of PPI scoring functions. 

We therefore uniformly sample in DockQ the models generated from rigid-body redocking of the main dataset of 84 heterodimers and score the resulting models with each of the six scoring functions (ZRank2, ITScorePP, Rosetta, PyDock, VoroMQA, and Deeprank-GNN-ESM). In Fig. \ref{fig:per_target_spearman_scores}, we show the Spearman correlation coefficient $\rho$ between ${\rm DockQ}$ and six scoring functions for the rigid-body re-docked models from the $84$ PPI targets. The targets are ordered according to increasing $\langle \rho \rangle$, averaged over the six scoring functions. (Note that since $\rho < 0$, increasing $\rho$ implies decreasing magnitude of the correlations.) Some targets have $\rho \approx -1$, e.g. 3WHQ, 1UGQ, and 1N1J. Approximately $20$ targets are ``easy'' to score with $|\langle \rho \rangle| > 0.8$. Twenty-three targets are ``hard'' to score with $|\langle \rho \rangle| < 0.6$. The remaining $41$ targets with $0.8 \leq |\langle \rho \rangle | \leq 0.6$ are medium difficulty. Medium and hard targets display higher variance in $|\langle \rho \rangle|$ than easy targets. Also, some of the scoring functions, like Deeprank-GNN-ESM and VoroMQA, underperform (with smaller $|\langle \rho \rangle|$) on medium and hard targets, suggesting they are less effective at scoring models for these targets. We show the average $\langle \rho \rangle_t$ over targets for each scoring function in Fig. \ref{fig:simple_spearman_bm55} in Appendix \ref{app:a}. ZRank2 has the largest $|\langle \rho \rangle _t| \approx 0.78$. In contrast, VoroMQA has the smallest $|\langle \rho \rangle _t| \approx 0.56$ (excluding GNN-DOVE) and performed well only on the easy targets. We carried out the same analysis of scoring functions on an additional $62$ heterodimer targets from the ZDOCK Benchmark-5.5 dataset \cite{vreven_updates_2015} and found that the relative performance of the scoring functions was nearly identical to that in Fig.~\ref{fig:simple_spearman_bm55}. (See Appendix \ref{app:a} for more information.) These results show that current scoring functions give accurate predictions (with $|\rho| > 0.8$) of model quality for only $\sim 25 \%$ of the PPI targets with high-resolution x-ray crystal structures. A highly accurate scoring function would generate correlations $|\rho| \geq 0.9$ between the score and ${\rm DockQ}$ for all PPI targets.

\subsection{Physical features of protein interfaces determine their scoring difficulty}
\label{subsec:results_physical_features}

% Figure 5
\begin{figure*}[htbp]
\centering
\includegraphics[width=0.85\textwidth]{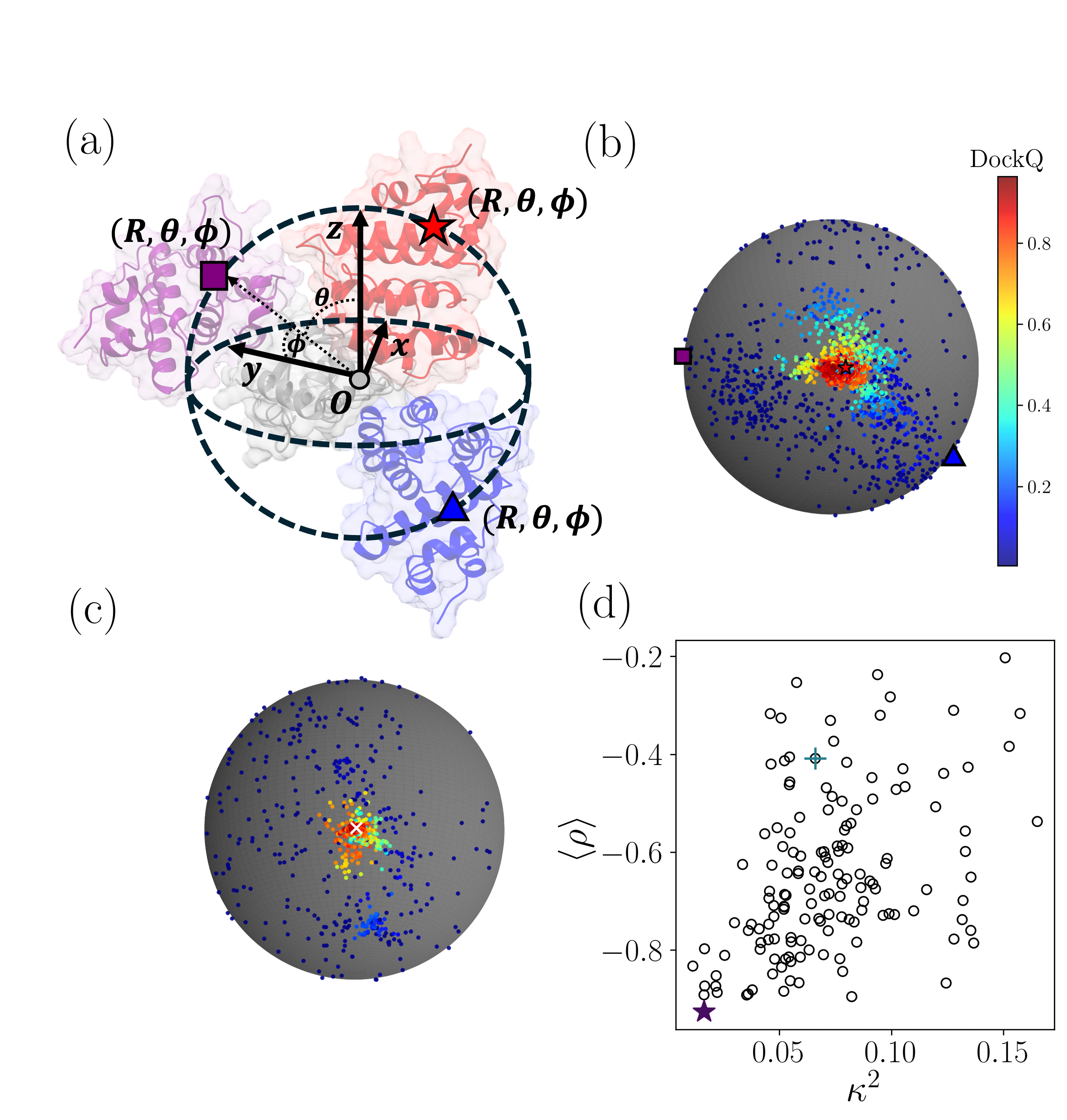}
\caption{The shape of the DockQ landscape provides insight into the effectiveness of scoring computational models. (a) A schematic of rigid-body re-docked models of heterodimer PDB 2GRN, which illustrates the variables that define the DockQ landscape. The receptor (gray shading) is located at the origin $O$ with the same orientation as that in the x-ray crystal structure. The location of the ligand with the same orientation as that in the x-ray crystal structure is denoted using spherical coordinates, the distance from the origin $r$, the polar angle $\theta$, and the azimuthal angle $\phi$, for two models (blue triangle and purple square) and the x-ray crystal structure (red star). DockQ increasing from dark blue to dark red is plotted for all computational models for (b) 2GRN and (c) 3WHQ with Spearman correlations between DockQ and ZRank2 $\langle \rho \rangle = -0.41$ and $-0.93$, respectively. The models were scored at arbitrary $x$-, $y$-, and $z$-coordinates, but plotted at $r=R$, where $R$ is the separation between the centers of mass of the receptor and ligand in the x-ray crystal structure. The white ``x'' denotes the location of the ligand in the x-ray crystal structure for 3WHQ. (d) The Spearman correlation $\langle \rho\rangle$ (between DockQ and PPI score) averaged over scoring functions plotted versus the relative anisotropy $\kappa^2$ of the DockQ landscape for all $84$ targets. The values of $\rho$ and $\kappa^2$ are highlighted for 2GRN (cross) and 3WHQ (star).}
\label{fig:bullseye_analysis}
\end{figure*}

What makes a PPI target easy or hard to score? Rigid-body re-docking involves generating models in a six-dimensional configuration space. If the receptor's center of mass is fixed at the origin, the ligand's center of mass can be placed at any position $(x,y,z)$ (three degrees of freedom) and three angles are required to specify its orientation. We can calculate the PPI score and DockQ score at any point in this six-dimensional space, and a large $|\rho|$ indicates strong correlations between the PPI score and DockQ in this six-dimensional configuration space. However, it is difficult to visualize and characterize the DockQ and PPI score landscapes in six dimensions, and thus we seek lower-dimensional representations. One approach is to fix the orientation of the ligand (e.g. to that in the x-ray crystal structure) and only consider the $x$-, $y$-, and $z$-coordinates of the ligand's center of mass (given a fixed position and orientation of the receptor). We can remove another degree of freedom by plotting the scores at fixed radial distance $r=R$, where $R$ is the distance between the receptor and ligand centers of mass in the x-ray crystal structure. We can then visualize the DockQ score (and the PPI score) for each model as function of the spherical coordinate representation of the ligand's center of mass position ($R,\theta,\phi$), where $\theta$ and $\phi$ are the polar and azimuthal angles (Fig. \ref{fig:bullseye_analysis} (a)). 

We show the DockQ landscape for all computational models for the targets PDB 2GRN and 3WHQ in Fig. \ref{fig:bullseye_analysis} (b) and (c). The models with the largest DockQ values in red are close to the native binding sites. The DockQ score decays rapidly (changing color from red to yellow to dark blue) as the models move further from the native binding site in $\theta$ and $\phi$. As shown in Fig. \ref{fig:per_target_spearman_scores}, 2GRN is classified as a ``hard'' target  with $\langle \rho \rangle = -0.41$ and 3WHQ is classified as an ``easy'' target with $\langle \rho \rangle = -0.93$. The DockQ landscape is more isotropic for the easy target 3WHQ compared to the landscape for the hard target 2GRN. To characterize the shape of the DockQ landscape, we calculate the moment of inertia tensor $I_{\alpha \beta}$ for the $x$-, $y$- and $z$-coordinates of the ligands for all models weighted by DockQ for each target. We define the relative anisotropy, $\kappa^2 = 1-3(\lambda_1 \lambda_2 + \lambda_2 \lambda_3 + \lambda_1 \lambda_3)/(\lambda_1 + \lambda_2 + \lambda_3)^2$, of the landscape, where $\lambda_1 \ge \lambda_2 \ge \lambda_3$ are the eigenvalues of $I_{\alpha\beta}$. (See Sec.~\ref{assess} for more details.) We show in Fig. \ref{fig:bullseye_analysis} (d) that there is a positive correlation between $\kappa^2$ and $\langle \rho \rangle$ averaged over all scoring functions, i.e. $\langle \rho \rangle$ decreases (becomes more negative) as the large-DockQ landscape becomes more isotropic. 

The magnitude of the Spearman correlation between DockQ and the PPI score is controlled by the differences in the DockQ and PPI score landscapes. In Fig.~\ref{fig:app_easy_hard_sphere} of Appendix~\ref{app:d}, we compare the DockQ and ZRank landscapes of PDB 3WHQ and 2GRN. For the easy target, 3WHQ, the ZRank landscape closely resembles the DockQ landscape, while the ZRank landscape does not overlap with the DockQ landscape for the hard target, 2GRN. These results emphasize that the isotropy of the DockQ landscape and the overlap between the DockQ and PPI score landscape influence $\langle \rho \rangle$ for each target. However, are there {\it physical} features of the binding interface that can provide information about the difficulty in accurately scoring PPI models? 

\begin{figure}[htbp]
\centering
\includegraphics[width=0.95\columnwidth]{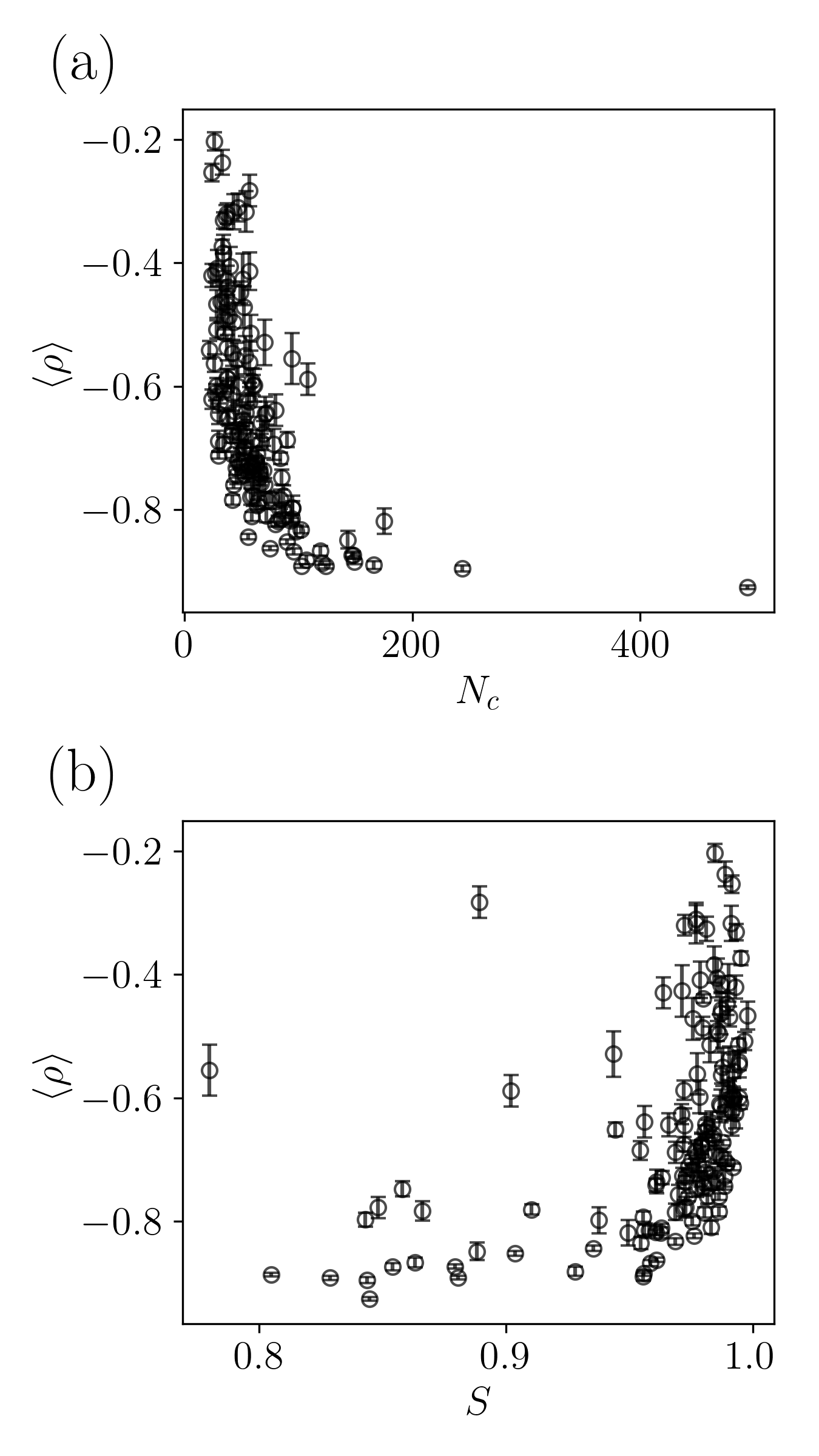}
\caption{The Spearman correlation $\langle \rho \rangle$ between the DockQ and PPI scores, averaged over all scoring functions, plotted versus (a) the number of interfacial contacts $N_c$ and (b) the interface separability $S$. The error bars show the standard error of the mean.}
\label{fig:flatness_and_rel_int}
\end{figure}

We identify two important physical properties of protein-protein interfaces that influence scoring accuracy: the interface separability $S$ and the number of interfacial contacts between heavy atoms $N_c$. $N_c$ was chosen as a feature because of its use in many PPI scoring functions \cite{dominguez_haddock_2003,chen_zdock_2003, gray_proteinprotein_2003, cheng_pydock_2007, pierce_zrank_2007, pierce_combination_2008,vreven_updates_2015, olechnovic_voromqa_2017, marze_efficient_2018}. Scoring functions should maximize $N_c$, since computational models with higher values of $N_c$ are typically closer to the x-ray crystal structure of the target (i.e. possess larger DockQ). $N_c$ is calculated by counting all interfacial contacts in the target between heavy atoms at the interface with separations $\leq 4.5$ \AA. For each target, as $N_c$ increases, $\langle \rho \rangle$ becomes more negative, indicating that it is easier to score models of targets with larger $N_c$. (See Fig. \ref{fig:flatness_and_rel_int} (a).)
 
The interface separability $S$ is calculated using a support vector machine (SVM) to best separate the receptor and ligand by an order-three polynomial surface. Thus, $S$ is a measure of the geometric complementarity of the protein-protein interface \cite{ chen_novel_2003, berchanski_hydrophobic_2004, liu_combinatorial_2006, geppert_protein-protein_2010, mccafferty_simplified_2021}. The separability satisfies $0.5 \le S \le 1.0$ and the larger the separability, the flatter the interface. In Fig. \ref{fig:flatness_and_rel_int} (b), we show that as the interfaces become more intertwined ($S$ decreases), $|\langle \rho \rangle |$ averaged over scoring functions increases. In contrast, as $S \rightarrow 1$, $| \langle \rho \rangle |$ generally decreases, but the data for different targets are more scattered.

Since ``easy'' targets have a more isotropic DockQ landscape, and easy targets have low values of the separability and high values of $N_c$, the anisotropy $\kappa^2$ is typically smaller for targets with low $S$ and large $N_c$ as shown in Fig. \ref{fig:supp_anisotropy_vs_physical} in Appendix \ref{app:d}. These results show that the physical properties of the interface, such as the separability and number of contacts, enable us to distinguish the ``easy'' from the ``hard'' targets when scoring PPI models. We also show in Fig. \ref{fig:bm55_comparison} in Appendix \ref{app:d} that the distributions of $N_c$ and $S$ for the main $84$-target dataset are similar to those for the ZDOCK Benchmark-5.5 dataset~\cite{vreven_updates_2015}.

% Figure 6

\subsection{A two-feature SVR model can match the accuracy of current scoring functions}
\label{subsec:results_svr}

\begin{figure*}[htbp]
\centering
\includegraphics[width=0.9\textwidth]{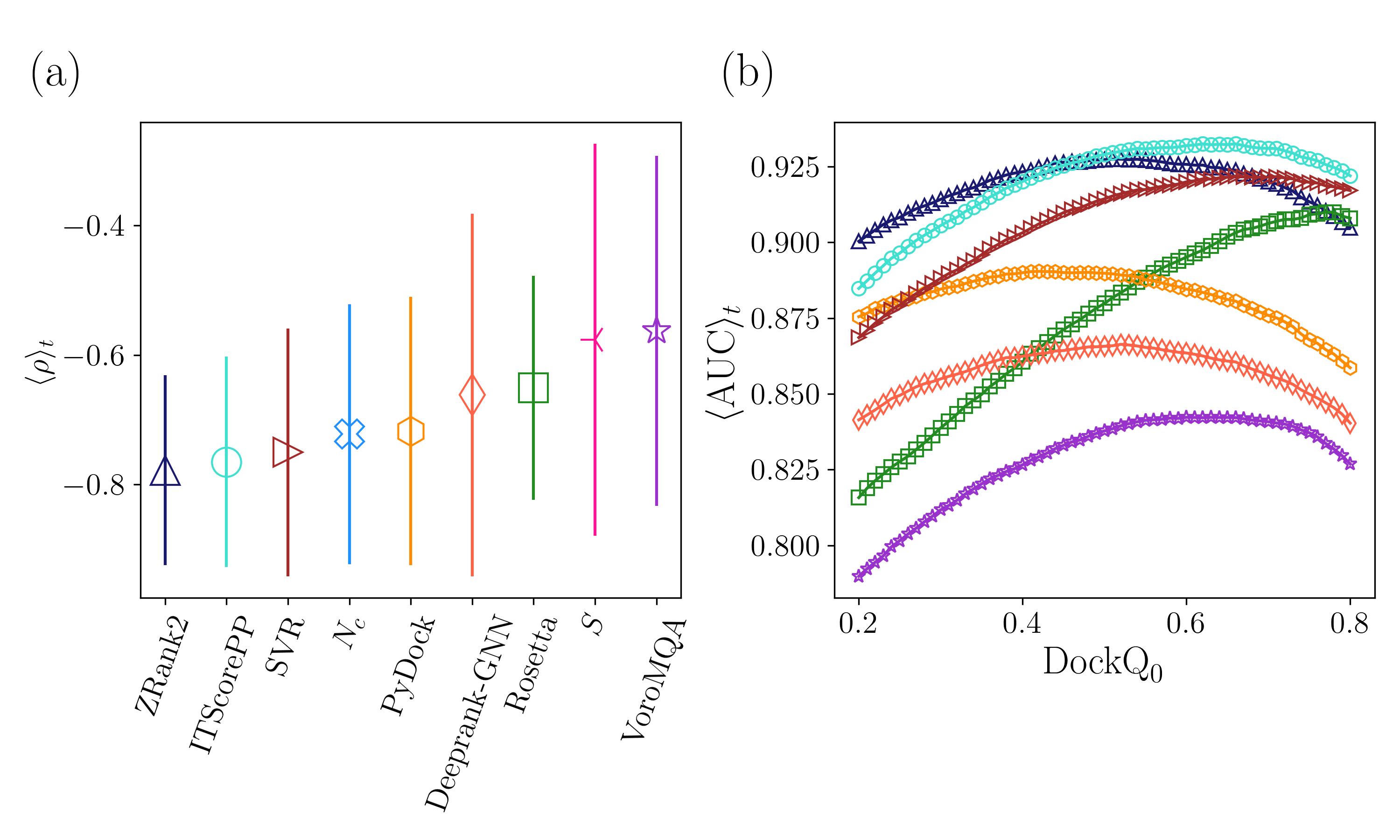}
\caption{(a) The Spearman correlation between DockQ and the PPI score $\langle \rho \rangle_t$ (markers) averaged over targets and the standard deviation $\langle \Delta \rho \rangle_t$ (vertical lines) for each scoring function (including the individual physical features: $N_c$ and $S$). $\langle \rho \rangle_t$ for each scoring function is plotted in ascending order from left to right. (b) The AUC of the ROC plot $\langle$AUC$\rangle_t$ averaged over targets plotted versus the DockQ cutoff, ${\rm DockQ}_0$, for each scoring function: ZRank2 (blue triangles), ITScorePP (turquoise circles), Rosetta (green squares), PyDock (orange hexagons), VoroMQA (purple stars), Deeprank-GNN-ESM (red-orange diamonds), and the SVR scoring function with two physical features (burgundy rightward triangles).}
\label{fig:auc_and_physical_plot}
\end{figure*}

The physical interface properties, such as $S$ and $N_c$, can separate targets according to their scoring difficulty. Therefore, it is interesting to determine how these two interface properties perform as scoring functions for computational models. We first calculate these physical properties on all models for each target that are uniformly sampled over DockQ. We then plot the Spearman correlation between DockQ and the PPI scores $\langle \rho \rangle_t$ averaged over all targets for each scoring function, including scores that are based on $S$ and $N_c$ individually (Fig.~\ref{fig:auc_and_physical_plot} (a)). The individual scores based on the interface separability and number of contacts yield average Spearman correlation coefficients,  $\langle \rho \rangle_t=-0.577$ and $-0.722$, respectively, showing that $N_c$ is a more effective score than $S$. 

The success of the two physical interface properties individually in scoring computational models suggests that a combined scoring function that includes both physical metrics can outperform them individually. In Fig \ref{fig:auc_and_physical_plot}, we show results for a non-linear support vector regression (SVR) score that uses a gaussian kernel to combine the two physical features. For training and testing the SVR, we remove six targets from the dataset that contain monomers with similar sequences to others in the dataset. There are $10$ targets in the $84$ target dataset with one monomer that has high sequence similarity ($\geq 90\%$) to another in the dataset: 1WRD, 2HTH, 3K9P, 7JXV, 1J7D, 4DHI, 2GRN, 2PE6, 1PDK, and 2WMP. To avoid potential bias during the training and testing of the SVR score, we removed six of the ten structures to ensure no heterodimers in the dataset shared monomers with high sequence similarity. We kept four targets (3K9P, 4DHI, 2GRN, and 2WMP) because they had the lowest $|\langle \rho \rangle|$ out of the structures with which they shared a monomer. Thus, we trained and tested the SVR only on $78$ targets in the main $84$ target dataset. The SVR score achieves $|\langle \rho \rangle_t| = 0.750$, which is only slightly lower than the highest performing PPI score ZRank2 with $|\langle \rho \rangle_t| = 0.778$. Thus, this simple two-feature SVR scoring function can match the performance of state-of-the-art PPI scoring functions. 

We then calculate the classification performance for the PPI scoring functions and the combined SVR score using the AUC of the ROC curve. Because the ROC requires an arbitrary cutoff between positives and negatives, we calculate AUC over a range of DockQ cutoffs $0.2 \leq {\rm DockQ}_0 \leq 0.8$. If the DockQ score of a model is greater than or equal to the cutoff, it is classified as positive.  In Fig. \ref{fig:auc_and_physical_plot} (b), we plot $\langle {\rm AUC} \rangle_t$ averaged over targets versus the cutoff ${\rm DockQ}_0$. The ordering of the Spearman correlation mirrors the classification performance at the lowest cutoff (${\rm DockQ}_0 = 0.2$) in Fig. \ref{fig:auc_and_physical_plot} (b). As the DockQ cutoff increases, the classification performance of most scoring functions does not change significantly. However, $\langle {\rm AUC} \rangle_t$ for Rosetta increases by more than $0.1$ as DockQ$_0$ increases, which suggests that Rosetta is better at identifying high-quality models. Notably, when $\rm{DockQ}_0 = 0.8$, ITScorePP and the SVR score are the top classifiers of high-ranking models, both with $\langle$AUC$\rangle_t \approx 0.92$. We further tested the SVR model trained on the $78$ targets in the main PPI target dataset on models generated using the $62$ heterodimer targets from the ZDOCK Benchmark-5.5 dataset. Again, we find that the SVR model had the third best performance of the scoring functions tested. These results emphasize that the correlation and classification performance of the two-feature SVR model is comparable to that for current PPI scoring functions. 

% Figure 7

\subsection{Scoring effectiveness of docked models decreases as iRMSD from the bound conformation increases}
\label{subsec:flexible}

\begin{figure*}[htbp]
    \centering
    \includegraphics[width=1\linewidth]{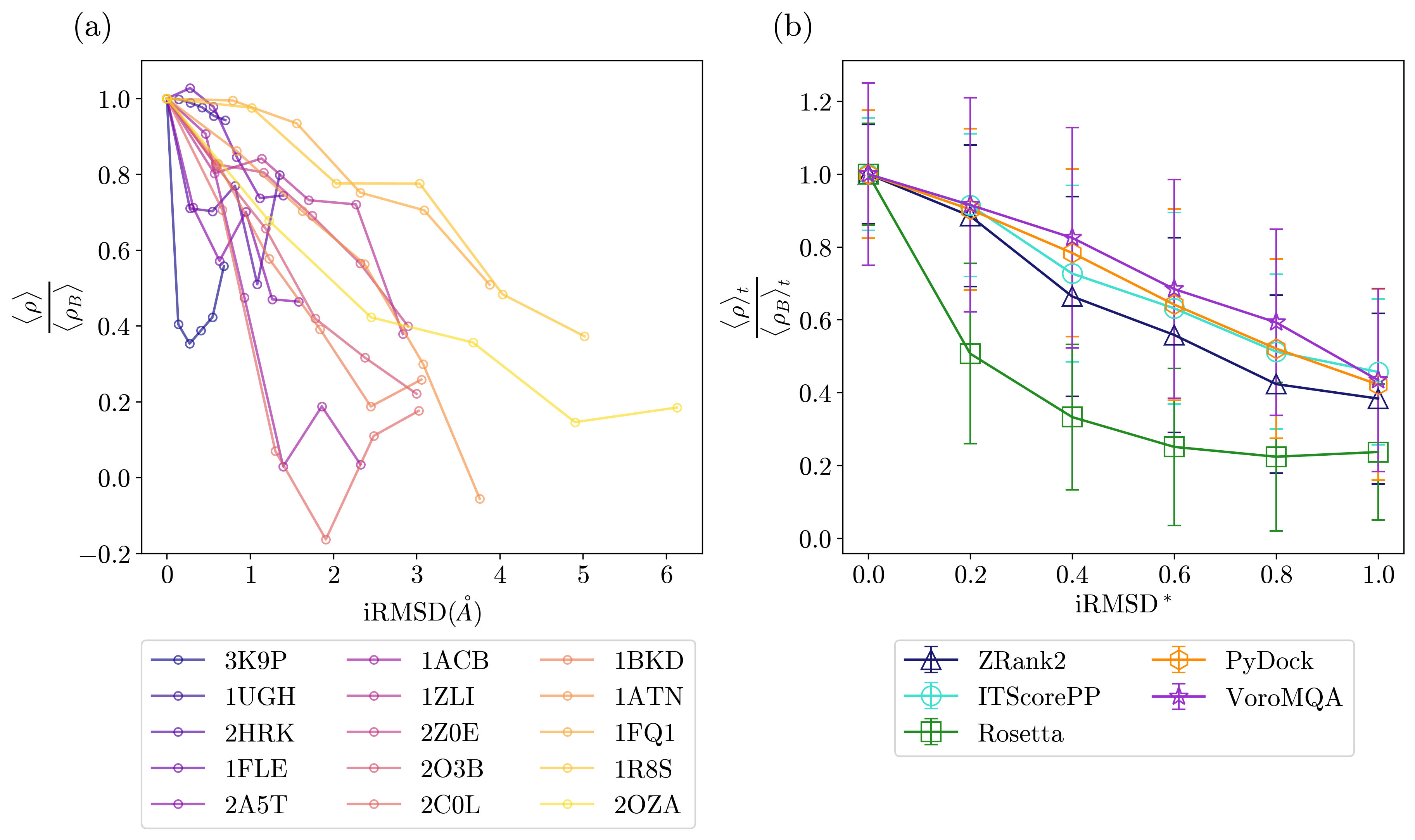}
    \caption{Performance of PPI scoring functions on docked monomers as a function of ${\rm iRMSD}$. (a) The Spearman correlation $\langle\rho\rangle$ between DockQ and PPI score averaged over the five scoring functions for $15$ PPI targets plotted versus the ${\rm iRMSD}$ of the interface C$_\alpha$ atoms for monomers that are deformed from their bound to unbound conformations. The average correlation is normalized by the average Spearman correlation $\langle\rho_B\rangle$ from scoring rigid-body re-docked models of the bound monomer conformations. The largest ${\rm iRMSD}$ data point for each target represents docking of the unbound monomer conformations.  (b) The Spearman correlation $\langle\rho\rangle_t$ for each PPI scoring function averaged over the fifteen targets (normalized by the average $\langle\rho_B\rangle_t$ from scoring rigid-body re-docked models of the bound monomer conformations) plotted versus the normalized ${\rm iRMSD}^*$.  The error bars represent the standard deviation of the Spearman correlation for each scoring function across the $15$ targets.}
    \label{fig:interpolation}
\end{figure*}

We have demonstrated significant variability of the scoring function performance even in the limit of rigid-body re-docking, where the RMSD between the bound and unbound monomer conformations is zero. However, monomers can undergo large conformational changes as they bind. In this section, we investigate how changes in the monomer conformations from their bound forms affects the Spearman correlation $\rho$ between the PPI score and DockQ for each of the five scoring functions (ZRank2, ITScorePP, Rosetta, PyDock, and VoroMQA). For this study, we selected heterodimers whose monomers have high-resolution x-ray crystal structures deposited in the PDB in their unbound forms. We chose thirteen heterodimers from the ZDOCK Benchmark-5.5 dataset and two heterodimers from the main dataset of $84$ heterodimers, as listed in Fig. \ref{fig:interpolation} (a). Of the $15$ heterodimer targets, the ${\rm iRMSD}$ between the bound and unbound monomers ranges from 0.68\AA\ (PDB 3K9P) to 6.13\AA\ (PDB 2OZA). Since the $13$ heterodimer targets from the ZDOCK Benchmark-5.5 dataset appear in the Deeprank-GNN-ESM training set, Deeprank-GNN-ESM was not included in this assessment~\cite{xu_deeprank-gnn-esm_2024}. As discussed in Sec.~\ref{subsec:sampling}, we linearly interpolate the conformations of the monomers between their bound and unbound forms to generate a series of six images that span the full range of ${\rm iRMSD}$. For each image, we dock the pairs of monomers using uniform sampling in DockQ, score the models using the five scoring functions, and calculate the Spearman correlation between DockQ and each scoring function. 

In Fig.~\ref{fig:interpolation} (a), we show the Spearman correlation for each target $\langle\rho\rangle$ averaged over the five scoring functions normalized by the average Spearman correlation $\langle\rho_B\rangle$ for the rigid-body re-docked models as a function of ${\rm iRMSD}$. While there are some variations,  $\langle\rho\rangle$ decreases strongly with ${\rm iRMSD}$ for all targets. In particular, when ${{\rm iRMSD}} = 1.5~\mathring{\text{A}}$, i.e. the cutoff for the easiest ``rigid-body'' docking category in the ZDOCK Benchmark-5.5 dataset, $9$ out of the $15$ heterodimer targets have $\langle\rho\rangle/\langle\rho_B\rangle\le0.8$. We also analyze the performance of each scoring function after averaging the Spearman correlation over the $15$ targets to obtain $\langle\rho\rangle_t$. In Fig.~\ref{fig:interpolation} (b), we plot $\langle\rho\rangle_t$ normalized by the Spearman correlation $\langle \rho_B\rangle_t$ for models from rigid-body re-docking as a function of the rescaled ${\rm iRMSD}^*={\rm iRMSD}/{\rm iRMSD}_U$, where ${\rm iRMSD}_U$ is the iRMSD between the bound and unbound monomer conformations. For all scoring functions, $\langle\rho\rangle_t/\langle\rho_B\rangle_t$ decreases monotonically with ${\rm iRMSD}^*$. We find that the average Spearman correlation ranges from $\langle\rho\rangle_t/\langle\rho_B\rangle_t=0.24$ (Rosetta) to $0.46$ (ITScorePP) at ${\rm iRMSD}^*=1$. This rapid drop in Spearman correlation indicates that current scoring functions are not able to accurately assess model quality during flexible docking.

% Figure 8

\section{Conclusions and Future Directions}
\label{discussion}

Predicting the interface between two protein monomers that form a heterodimer complex is an important, open problem. In this work, we focus on perhaps the simplest aspect of this problem: rigid-body re-docking of the bound forms of the monomers into the heterodimer, for which we have an experimentally-determined structure. When re-docking the bound forms, generation of the computational models is straightforward. Instead, the difficult task is to identify the model that is closest to the native structure using a PPI scoring function. We evaluate the performance of seven PPI scoring functions (ZRank2, PyDock, VoroMQA, Rosetta, ITScorePP, GNN-DOVE, and Deeprank-GNN-ESM) by determining the Spearman correlation between the score and DockQ for the computational models. We test these scoring functions on models from a main dataset of $84$ heterodimer targets, for which we have high-resolution x-ray crystal structures of the individual monomers and dimers. We validate our results on the ZDOCK Benchmark-5.5 dataset of $64$ non-redundant heterodimer targets. For each target, we generated an equal number of positive (DockQ $\geq 0.23$) and negative (DockQ $< 0.23$) models. The DockQ-balanced set of models gives consistent results for both classification and correlation metrics, i.e. we find that the Spearman correlations $\rho$ between the PPI score and DockQ and the $\rm AUC$ of the ROC curve obey $\rm AUC \approx -0.5 \rho + 0.5$. We further show that the targets in our datasets possess a range of difficulty in scoring, where some targets have significantly lower $|\rho|$ than others. We show that the scoring difficulty of the target increases as the anisotropy of the DockQ scoring landscape increases. Further, we identified several physical properties of the interface, such as interface separability $S$ and the number of interfacial contacts $N_c$ that can predict the scoring difficulty of the targets. We then constructed an SVR scoring function based on $S$ and $N_c$ and showed that it meets or exceeds the classification and correlation performance of the other PPI scoring functions.

We found that an SVR model with only two physical features matches or outperforms state-of-the-art PPI scoring functions with many more features, despite the long history of computational studies of PPIs~\cite{wodak_computer_1978, shoichet_protein_1991, cherfils_protein-protein_1991, camacho_scoring_2000, pierce_zrank_2007, cheng_pydock_2007, olechnovic_voromqa_2017, olechnovic_voroif-gnn_2023}. One possible reason for this lagging performance for current PPI scoring functions~\cite{pierce_zrank_2007, cheng_pydock_2007, pierce_combination_2008, olechnovic_voromqa_2017, pang_deeprank_2017, wang_protein_2020, wang_protein_2021, reau_deeprank-gnn_2023, xu_deeprank-gnn-esm_2024} is that they have primarily been evaluated using classification-based metrics, such as hit rate plots, which employ quality cutoffs to define positive or negative models. However, using low cutoffs will make PPI scoring functions seem  ``successful,'' even if the scoring functions possess very weak correlations with DockQ. Classification performance can be informative, yet it should not be the sole metric upon which the PPI scoring function performance is assessed. 

Why do we focus on rigid-body re-docking of bound monomers, rather than starting from unbound monomers and tackling the more general problem of flexible docking? Addressing this question would require a scoring function that remains highly accurate and strongly correlated with DockQ even when the monomers are far from their bound conformations (i.e., ${\rm iRMSD} \gtrsim 1.5$~\AA). However, as shown in Fig.~\ref{fig:interpolation}, the effectiveness of all scoring functions considered here—quantified by the Spearman correlation between the score and DockQ—decreases strongly for ${\rm iRMSD} > 0$. This rapid decrease in correlation highlights a fundamental limitation in assessing scoring function performance for flexible docking. Moreover, in the absence of experimentally resolved structures that interpolate between the bound and unbound monomer conformations, there is no clear ground truth against which to validate the computational models generated during flexible docking. Together, these points underscore why rigid-body re-docking of bound monomer conformations provides a necessary first step for evaluating PPI scoring functions and solving the more general flexible docking problem.

We have shown that the recent supervised GNN-based scoring functions, GNN-DOVE \cite{wang_protein_2021} and Deeprank-GNN-ESM \cite{xu_deeprank-gnn-esm_2024}, have worse performance for the Spearman correlation between score and DockQ than all other scoring functions that we tested. One possible reason for the worse performance of these supervised GNN-based scoring functions is the quality imbalance in the datasets used for training. For example, GNN-DOVE is trained on the DockGround1.0 dataset \cite{kundrotas_dockground_2018}, where only $\sim 9\%$ of models have $\rm DockQ \geq 0.23$. Deeprank-GNN-ESM created their own training set, yet only $\sim 11\%$ of models have $\rm DockQ \geq 0.23$. Supervised learning methods are most effective when trained on datasets with a uniform distribution of model quality. It is likely that GNN-based scoring functions will have improved performance if they are re-trained on datasets with more uniform distributions of model quality \cite{mooijman_effects_2023}. Changing the architecture of the GNN to better represent physical features of the protein interface can also increase the scoring function performance. GNNs are composed of nodes (representing amino acids or atoms) with typically tens to thousands of features per node, yet they are currently unable to outperform the SVR in this study with only {\it two} physical features that describe the protein interface. In future studies, we will develop GNNs that first build-in informative physical features, such as $N_c$. We will add node and edge features to represent additional physical features of proteins, such as the hydrophobicity and electrostatic potential~\cite{papoian_role_2003, marze_modeling_2017, kopel_comparative_2019}. We aim to iteratively increase the number of features in GNN scoring functions and therefore identify the important physical features of protein interfaces that yield accurate scoring functions. In addition, we will employ structural clustering of models since it has been shown to reduce the complexity of sampling and scoring of PPI targets \cite{kozakov_optimal_2005, rodrigues_clustering_2012, kozakov_cluspro_2017}.

As discussed previously, template-free flexible docking is extremely difficult for PPI targets that undergo significant conformational changes upon binding. For example, recent studies were able to generate models where the interface C$_{\alpha}$ root-mean-square deviations (RMSD) relative to the x-ray crystal structure was less than $5$ \AA{} for only $31\%$ of the targets (i.e. $10$ out of $32$ targets that undergo significant conformational changes)~\cite{marze_efficient_2018}. To make more rapid progress in template-free flexible docking~\cite{torchala_swarmdock_2013, dominguez_haddock_2003, gray_proteinprotein_2003, marze_efficient_2018}, it is important to develop PPI scoring functions with Spearman correlations with DockQ of $|\rho| \approx 1$ for rigid-body re-docking datasets. Despite many years of development, all PPI scoring functions we tested have $|\langle \rho \rangle_t| < 0.8$ when averaged over all models and targets, but most of the scoring functions have $|\langle \rho \rangle_t| \leq 0.7$. Template-free flexible docking involves a series of rigid-body docking steps along a trajectory from the unbound to bound conformations of the monomers. We have shown that the correlations between current scoring furnctions and DockQ decrease rapidly with the interface RMSD of the monomers. Thus, since the PPI scores serve as the ``potential energy landscape'' that directs the system along the monomer trajectories from unbound to bound conformations, it is difficult to make progress in computational modeling of flexible docking of PPIs. 

Alphafold $3$ is frequently employed to generate heterodimer models of at least ``acceptable'' CAPRI quality~\cite{abramson_accurate_2024}, yet more work is needed to understand its underlying scoring function. Alphafold $3$ reports that it can predict $76\%$ of bound structures of protein heterodimers with ${\rm DockQ} \geq 0.23$. However, these studies do not report correlations between the underlying scoring function (reported as iLDDT) and DockQ for a large sampling of the models for each target. Thus, in future work, we seek to determine the correlations between iLDDT and DockQ for a large set of Alphafold 3 models. These studies will be important for understanding the dynamics of protein-protein interactions as the monomers interact and deform into the bound structures. In addition, there are many related problems, such as predicting the binding location and affinity for protein-protein interactions {\it in vivo}, that Alphafold 3 has not yet solved due to the lack of all-atom {\it in vivo} structures. Thus, it is important to continue to develop physics-based approaches to predict PPIs, which are complimentary to deep-learning approaches.

\section*{Acknowledgments}

The authors acknowledge support from NIH Training Grant Nos. T32GM145452 (A. C., A. T. G., D. F., G. M., J. S., N. B., and C.S.O.) and 5T15LM007056-39 (A. C.). This work was supported by the High Performance Computing facilities operated by Yale's Center for Research Computing. 

\section*{Data Availability}

The code and dataset used to run the analyses and figures are available at this link: \url{ https://github.com/jakesumneryale/Protein_interface_assessment}.

\appendix
\renewcommand{\thesection}{\Alph{section}} % Sections: A, B, C...
\renewcommand{\thesubsection}{\thesection.\arabic{subsection}}

\section{Scoring performance for the ZDOCK benchmark-5.5 dataset}
\label{app:a}

The main text focused on calculations of the Spearman correlation between DockQ and the PPI scores for the main dataset of $84$ heterodimer targets in Table~\ref{tab:pdb_labels}. In this appendix, we test the generality of the results in the main text by calculating the Spearman correlation between DockQ and the PPI scores for the $62$ additional heterodimer targets in the ZDOCK benchmark $5.5$ dataset in Table~\ref{tab:bm55_labels} \cite{vreven_updates_2015}. We generated computational models uniformly sampled in DockQ for each of the new targets. We then scored all models for each target using ZRank2, ITScorePP, Rosetta, PyDock, and VoroMQA. We did not assess the performance of Deeprank-GNN-ESM for the ZDOCK benchmark $5.5$ dataset because Deeprank-GNN-ESM was trained on $60/62$ of the targets.

%% Table 1
\begin{table}[htbp]
    \centering
    \setlength{\tabcolsep}{12pt}
    \begin{tabular}{|c@{\hspace{12pt}}c | c@{\hspace{12pt}}c | c@{\hspace{12pt}}c|}
         \hline
         No. & PDB & No. & PDB & No. & PDB \\
         \hline\hline
         1  & 3WHQ & 29 & 3WDG & 57 & 1J7D \\
         2  & 1UGQ & 30 & 4H5S & 58 & 1WRD \\
         3  & 1N1J & 31 & 2HLA & 59 & 7JXV \\
         4  & 4UZZ & 32 & 1RY7 & 60 & 6XOD \\
         5  & 1KFU & 33 & 5UQ2 & 61 & 4DHI \\
         6  & 4G6T & 34 & 3GFK & 62 & 6KP3 \\
         7  & 5K7M & 35 & 6ZBK & 63 & 1XG2 \\
         8  & 1C3A & 36 & 4GI3 & 64 & 4QLP \\
         9  & 5BY8 & 37 & 7JOE & 65 & 3SGB \\
         10 & 1R8O & 38 & 3QC8 & 66 & 4HWI \\
         11 & 4CT6 & 39 & 4YII & 67 & 2HTH \\
         12 & 4QJF & 40 & 4DBG & 68 & 1ITB \\
         13 & 1VET & 41 & 7BXH & 69 & 6WG4 \\
         14 & 3VZ9 & 42 & 1ACB & 70 & 3K9P \\
         15 & 5MAW & 43 & 4U1C & 71 & 3FPN \\
         16 & 3DGP & 44 & 1EUV & 72 & 2DVW \\
         17 & 1PDK & 45 & 1AY7 & 73 & 3QHY \\
         18 & 1F60 & 46 & 1GGP & 74 & 3U82 \\
         19 & 2WMP & 47 & 3RCZ & 75 & 6FUD \\
         20 & 4JE3 & 48 & 4BI8 & 76 & 3HW2 \\
         21 & 1UGH & 49 & 1TFO & 77 & 1WMH \\
         22 & 2BKR & 50 & 6YXJ & 78 & 2HSM \\
         23 & 8ENF & 51 & 1QAV & 79 & 3ONA \\
         24 & 2OUL & 52 & 1WQJ & 80 & 3YGS \\
         25 & 2FHZ & 53 & 2PE6 & 81 & 2H7Z \\
         26 & 1SPP & 54 & 2G2U & 82 & 2GRN \\
         27 & 5DMB & 55 & 4K12 & 83 & 2D5R \\
         28 & 5MU7 & 56 & 5ZRZ & 84 & 1MBV \\
         \hline
    \end{tabular}
    \caption{PDB IDs of the x-ray crystal structures in the main dataset of heterodimer targets. The ``No.'', or number, column links each PDB ID to the order of the targets with increasing average Spearman correlation in Fig.~\ref{fig:per_target_spearman_scores}.}
    \label{tab:pdb_labels}
\end{table}

%% Table 2
\begin{table}[htbp]
    \centering
    \begin{tabular}{|c|c|c|}
        \hline
        PDB ID & PDB ID & PDB ID \\
        \hline\hline
        1E96 & 1ATN & 1AVX \\
        1BKD & 1BUH & 1BVN \\
        1CGI & 1CLV & 1D6R \\
        1DFJ & 1FC2 & 1FLE \\
        1FQ1 & 1GLA & 1H1V \\
        1HE8 & 1IRA & 1J2J \\
        1KAC & 1KTZ & 1KXP \\
        1M10 & 1MAH & 1NW9 \\
        1OC0 & 1OPH & 1PPE \\
        1R0R & 1R8S & 1T6B \\
        1TMQ & 1UDI & 1US7 \\
        1Y64 & 1Z5Y & 1ZHH \\
        1ZHI & 1ZLI & 2A5T \\
        2A9K & 2AYO & 2B42 \\
        2BTF & 2C0L & 2FJU \\
        2G77 & 2HLE & 2HRK \\
        2NZ8 & 2O3B & 2O8V \\
        2OOB & 2OT3 & 2OZA \\
        2SIC & 2SNI & 2UUY \\
        2VDB & 2Z0E & 3SGQ \\
        7CEI &  &  \\
        \hline
    \end{tabular}
    \caption{PDB IDs of the x-ray crystal structures for targets in the ZDOCK Benchmark-5.5 dataset.}
    \label{tab:bm55_labels}
\end{table}

%%Table 3
\begin{table*}
    \centering
    \begin{tabular}{|c |c  |c |c  |}\hline
         
         Complex PDB ID& Unbound Monomer 1 PDB ID& Unbound Monomer 2 PDB ID& ${\rm iRMSD}$ (\AA)\\\hline
\hline
         
         3K9P& 3E46\_A& 1AAR\_A& 0.68\\\hline
         1UGH& 1AKZ\_A& 2UGI\_A& 0.70\\\hline
         2HRK& 2HRA\_A& 2HQT\_A& 1.35\\\hline
         1FLE& 9EST\_A& 2REL\_A(4)& 1.40\\\hline
         2A5T& 1Y20\_A& 2A5S\_A& 1.58\\\hline
         1ACB& 2CGA\_B& 1EGL& 2.33\\\hline
         1ZLI& 1KWM\_A& 2JTO\_A(6)& 2.84\\\hline
         2Z0E& 2D1I\_A& 1V49\_A(1)& 2.90\\\hline
         2O3B& 1ZM8\_A& 1J57\_A& 3.00\\\hline
         2C0L& 1FCH\_A& 1C44\_A& 3.03\\\hline
         1BKD& 1CTQ\_A& 2II0\_A& 3.06\\\hline
         1ATN& 1IJJ\_B& 3DNI& 3.76\\\hline
         1FQ1& 1B39\_A& 1FPZ\_F& 3.88\\\hline
         1R8S& 1HUR\_A& 1R8M\_E& 5.02\\\hline
         2OZA& 3HEC\_A& 3FYK\_X& 6.13\\ \hline 
    \end{tabular}
    \caption{PDB IDs and the ${\rm iRMSD}$ of the experimentally resolved structures in the $15$ target dataset that was used to generate monomer conformations between the bound and unbound forms. The ``Complex PDB ID'' column contains the PDB ID of the heterodimer, while the ``Unbound Monomer 1 PDB ID'' and ``Unbound Monomer 2 PDB ID'' columns give the PDB IDs of the unbound monomers.  For monomer structures with more than one chain, underscores followed by a letter denote the chain ID that matches the given bound monomer in the complex. Likewise, a number enclosed by parentheses denotes the biological assembly number containing the specific monomer conformation in the PDB. The ``${\rm iRMSD}$'' column contains the ${\rm iRMSD}$ between the bound and unbound forms of the monomers as discussed in Sec.~\ref{subsec: irmsd}}.
    \label{tab:irmsd_pdbids}
\end{table*}

%% Figure 9

\begin{figure*}
    \centering
    \includegraphics[width=\textwidth]{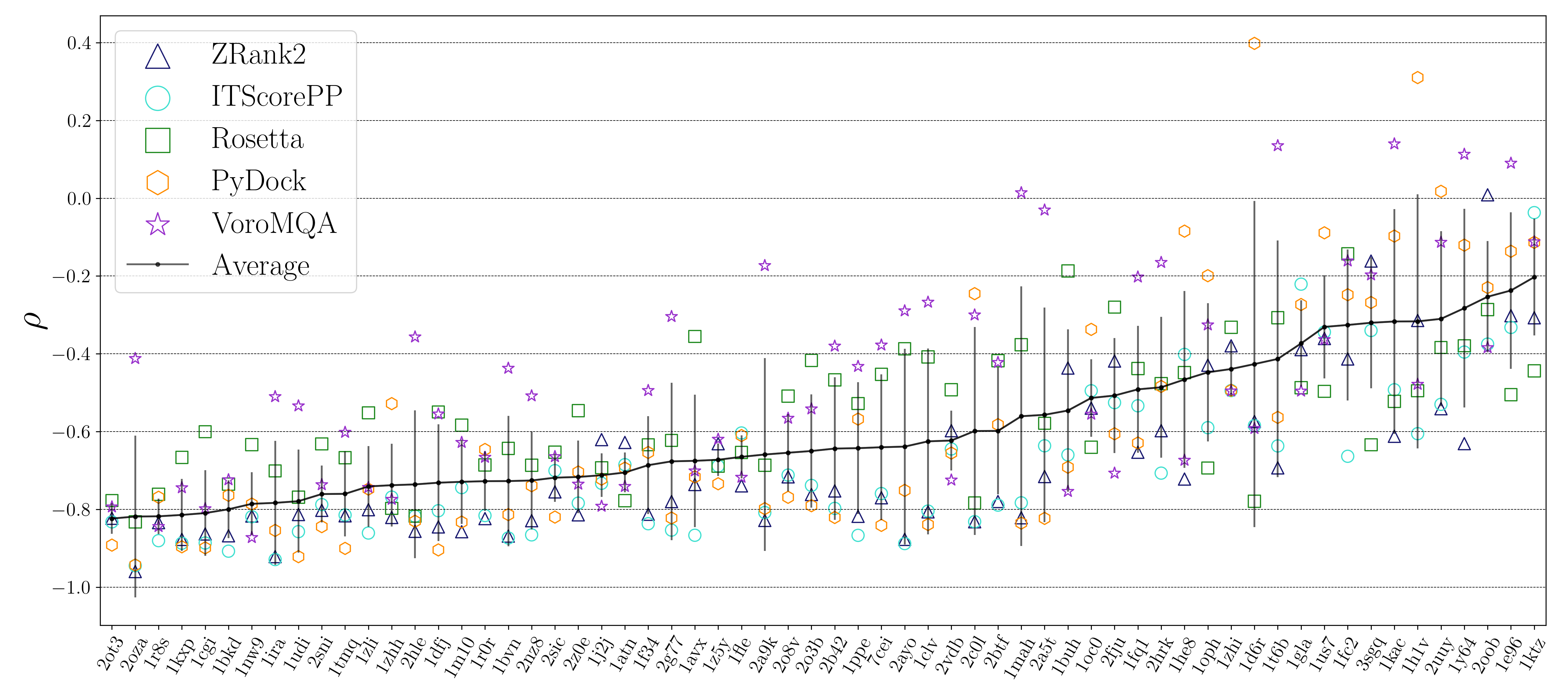}
    \caption{The Spearman correlation coefficient $\rho$ between DockQ and the PPI score for each of the $62$ targets in the ZDOCK Benchmark 5.5 dataset and scoring function. We order the targets by increasing $\langle \rho \rangle$ (black line), averaged over the scoring functions, from left to right. (Note that since $\rho < 0$ increasing $\rho$ implies decreasing magnitude of the correlations.) The standard deviations of the scores for each target are indicated by the black vertical lines. The PDB ID for each target is shown on the horizontal axis.}
    \label{fig:per_target_bm55}
\end{figure*}

We show $\rho$ for each of the targets in the ZDOCK benchmark-5.5 dataset in Fig.~\ref{fig:per_target_bm55}.  Based on $\langle \rho \rangle$, we find that this dataset contains a large fraction of medium- and hard to score targets, and only a few easy to score targets, i.e. $8.1\%$ are ``easy'',  $54.8\%$ are ``medium'', and $37.1\%$ are ``hard'' targets, where the thresholds are easy ($|\langle \rho \rangle| > 0.8$), medium ($0.6 \leq |\langle \rho \rangle| \leq 0.8$), and hard ($| \langle \rho \rangle| < 0.6$). In comparison, the original dataset of $84$ targets contained $23.8\%$ ``easy'',  $48.8\%$ ``medium'', and $27.4\%$ ``hard'' targets.

\begin{figure}
    \centering
    \includegraphics[width=\columnwidth]{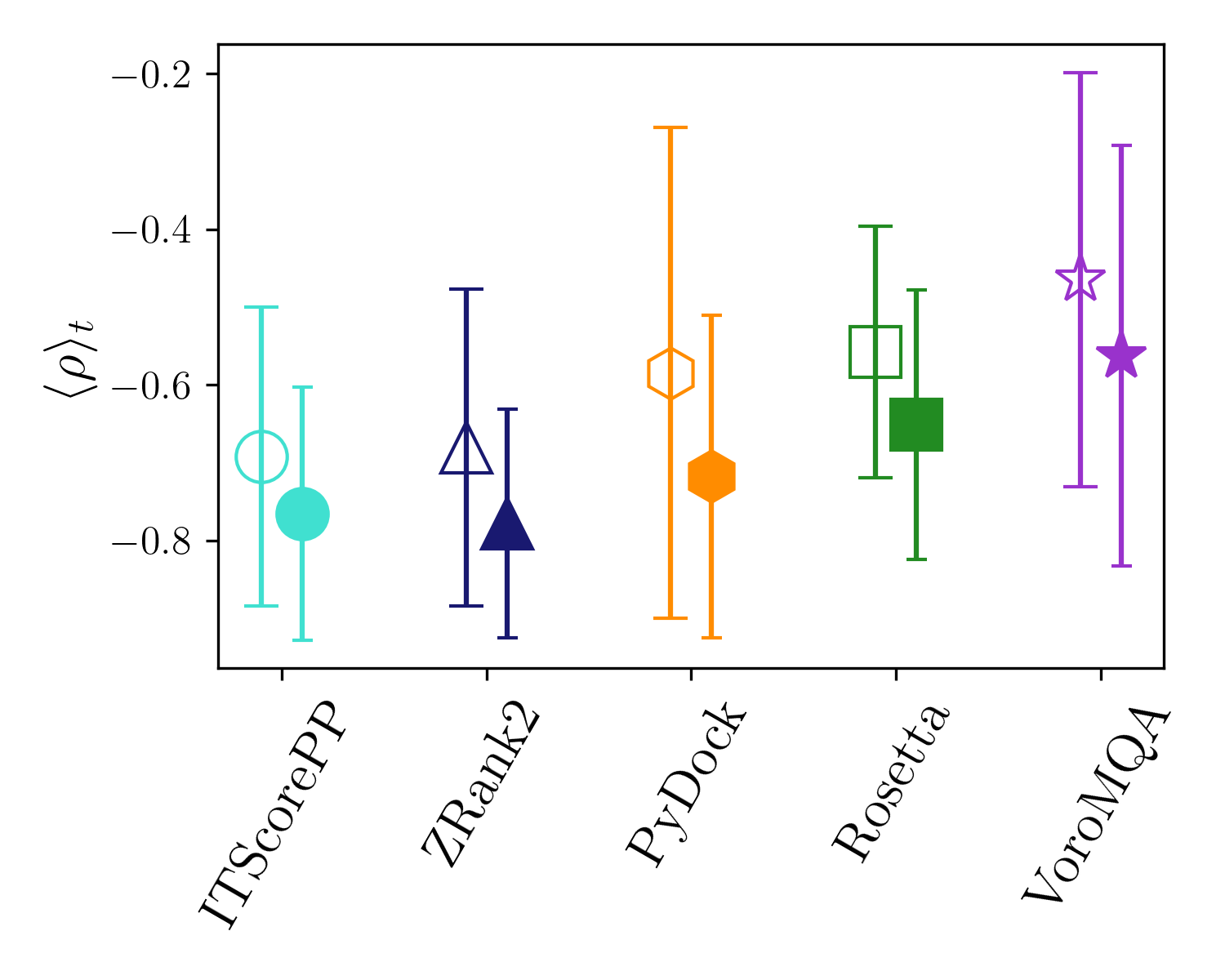}
    \caption{The Spearman correlation coefficient $\langle \rho \rangle_t$ averaged over targets in the ZDOCK Benchmark 5.5 dataset for each scoring function (open symbols) sorted in increasing order from left to right. The vertical lines represent $\pm \sigma$, were $\sigma$ is the standard deviation of $\rho$ across the targets for each scoring function. We also show $\langle \rho \rangle_t$ (filled symbols) for the main dataset of $84$ targets.}
    \label{fig:simple_spearman_bm55}
\end{figure}

We also report the Spearman correlation coefficient for each PPI score averaged over all targets $\langle \rho \rangle_t$ for both datasets in Fig.~\ref{fig:simple_spearman_bm55}. In general, the average Spearman correlation coefficient for each PPI scoring function is similar for each dataset. However, there are a few differences. The best performing PPI scoring functions for the ZDOCK benchmark-5.5 dataset are ITScorePP and ZRank2 with $\langle \rho \rangle_t \approx -0.7$. In contrast, the best performing scoring function is ZRank2 with $\langle \rho \rangle_t = -0.78$ for the main dataset with $84$ targets. In Fig.~\ref{fig:simple_spearman_bm55}, we show that all of the scoring functions have larger values of $|\langle \rho \rangle_t|$ for the main dataset of $84$ targets compared to the values for the ZDOCK benchmark-5.5 dataset. This result is expected since the $84$-target dataset possesses smaller values of $S$ compared to those for the ZDOCK benchmark-5.5 dataset. These results emphasize that although there are minor differences between the two datasets, the performance of the scoring functions is similar for both the original $84$ target and ZDOCK benchmark-5.5 datasets.

%% Figure 10

\newpage

\section{Protocols for calculating the scoring functions and DockQ}
\label{app:b}

The scoring and ground truth calculations (ZRank2 \cite{pierce_zrank_2007, pierce_combination_2008}, ITScorePP \cite{huang_iterative_2008}, Rosetta \cite{alford_rosetta_2017, marze_efficient_2018}, PyDock \cite{cheng_pydock_2007}, VoroMQA \cite{olechnovic_voromqa_2017}, Deeprank-GNN-ESM \cite{reau_deeprank-gnn_2023, xu_deeprank-gnn-esm_2024}, GNN-DOVE \cite{wang_protein_2021}, and ${\rm DockQ}$ \cite{basu_dockq_2016}) were performed using publicly available software. We also ran Rosetta Relax using the energy minimization protocols provided by Rosetta.

For the ZRank2 score, we downloaded the ZRank2 scoring function executable from the website \url{https://zdock.wenglab.org/software/download/#ZRANK}. The scoring function script is run by calling the executable with a text file listing the ``.pdb" files for which the program will calculate the ZRank2 scores. One ZRank2 score is provided per computational model for each dimer target. 

The ITScorePP executable was downloaded from
\url{http://huanglab.phys.hust.edu.cn/software/itscorepp}. From instructions for ITScorePP, we split each dimer into two separate monomer ``.pdb'' files and ran the executable with the flag ``-nomin'', which indicates that energy minimization should not be performed prior to scoring since we had already energy minimized the models using Rosetta Relax.  One ITScorePP score is provided per computational model.

The Rosetta software, including the \textit{Ref2015} scoring function was downloaded from \url{https://rosettacommons.org/software/download}. We calculated the \textit{Ref2015} Rosetta score for each computational model in the dimer form and for each monomer of the dimer separately. We then calculated the difference between the dimer Rosetta score and the sum of the monomer Rosetta scores to obtain the final Rosetta score, as recommended by Rosetta. This single Rosetta score is provided for each computational model for each dimer target. 

For the PyDock score, we used the scoring function that is included in the PyDock3 executable at \url{https://life.bsc.es/pid/pydock/get_pydock.html}. We created an input file that explicitly defines the receptor and ligand chains in the ``.pdb'' file for each computational model and passed the input file to the script with the flag ``bindEy'', which indicates that the executable should calculate the binding energy.  We identify the ``total score'' in the fourth column of the output as the PyDock score. 

For VoroMQA, we downloaded the ``voronota-voromqa'' executable from \url{https://github.com/kliment-olechnovic/voronota/blob/master/voronota-voromqa}. 
We ran the executable from the command line using the ``--score-inter-chain'' flag, which scores the interface residues. We identify the ``interaction score' as the VoroMQA score for each model.

For the Deeprank-GNN-ESM score, we downloaded the GNN model from \url{https://github.com/DeepRank/DeepRank-GNN-esm}. We ran the dowloaded python script to score each model from the command line with an input dimer ``.pdb'' file for each model. One Deeprank-GNN-ESM score is provided per model.

For the GNN-DOVE score, we downloaded the GNN model from \url{https://github.com/kiharalab/GNN_DOVE}. We ran the downloaded python script to score each model from the command line by specifying the input dimer ``.pdb'' file and the particular ``fold" (dataset that was used to train the GNN). We chose ``fold 5'', which was recommended by the authors. One GNN-DOVE score is given per model.

For the ${\rm DockQ}$ score, we downloaded the ${\rm DockQ}$ executable in Python using ``pip''. The source code for DockQ is available at \url{https://github.com/bjornwallner/DockQ}. We ran the ${\rm DockQ}$ executable from the command line for each model using the native dimer crystal structure as the reference. One ${\rm DockQ}$ score is output per model. 

We ran Rosetta Relax to energy minimize each computational model prior to scoring using Rosetta version 3.12, which can be downloaded at \url{https://rosettacommons.org/software/download}. We ran the command ``relax.static.linuxgccrelease''  with the flag ``-relax:contstrain\_relax\_to\_start\_coords'' to constrain the relaxation so each structure is only minimally perturbed. This relaxation protocol allows changes in the backbone and side chain dihedral angles, although the C$_{\alpha}$ root-mean-square displacement is nearly always less than $1.0$ \AA~from the starting structure.

\section{Sensitivity of hit rate plot on model quality distribution\label{app:c}}

%% Figure 11

\begin{figure*}[htbp]
\centering
\includegraphics[width=0.95\textwidth]{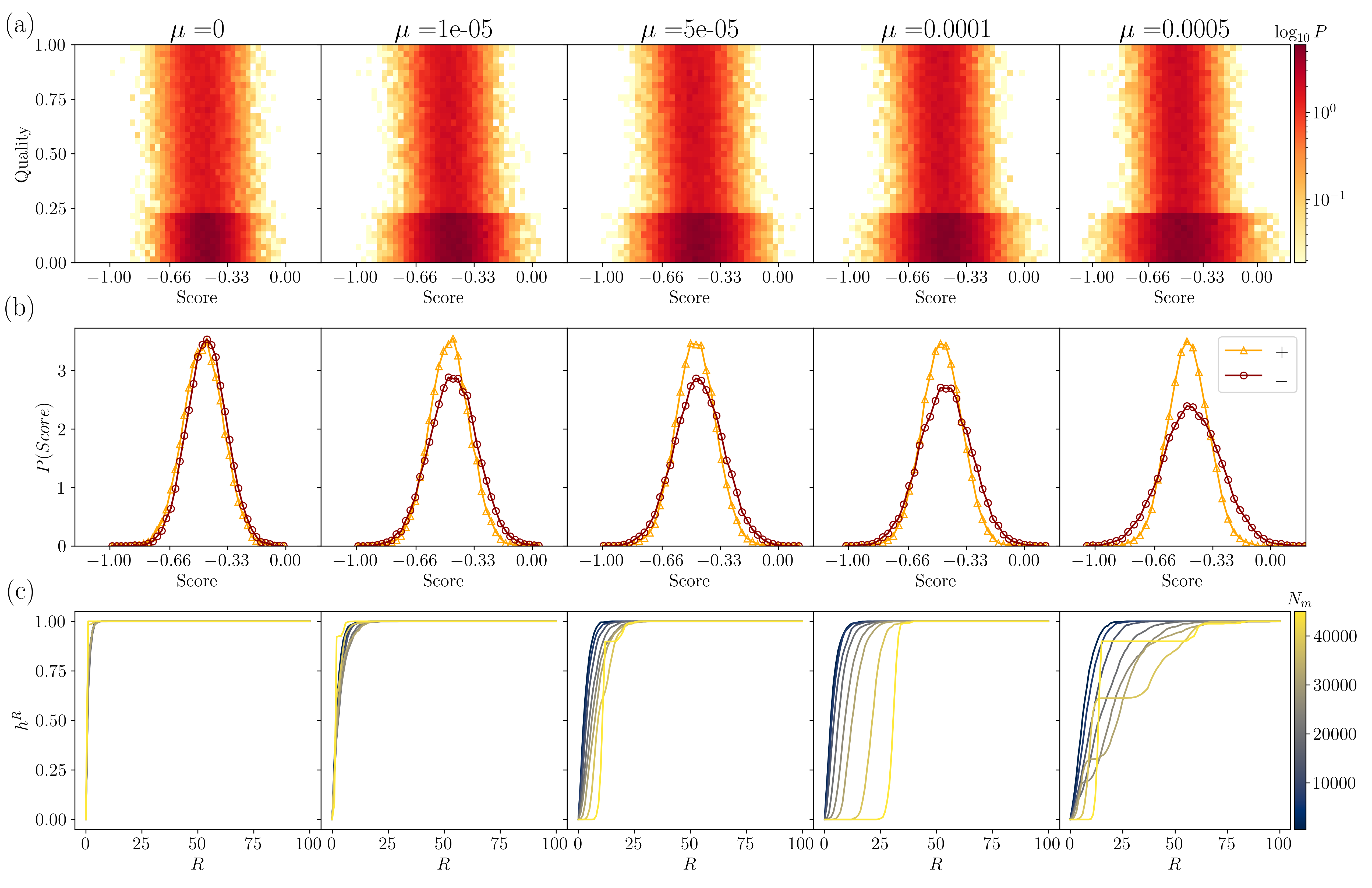}
\caption{The hit rate plot in the large number of models limit depends on the distribution of positive and negative models. (a) The joint probability distribution of quality and PPI score for $\sim$ 46,000 models, where the probability increases from white to dark red. The columns give different values of $\mu$, which is the fraction of computational models that are negative and  have a PPI score that is lower than the minimum value for all positive models. (b) The probability distributions of the PPI scores for positive (orange triangles) and negative (red circles) models for the data in (a). (c) The hit rate plots for the data in (a) plotted as a function of the number of models $N_m$.}
\label{fig:supp_synthetic_hr}
\end{figure*}

In Fig.~\ref{fig:supp_synthetic_hr}, we determine the sensitivity of the hit rate plot on the sample size and quality distribution of the models. We generated 108,000 data points with an initial Spearman correlation $\rho=-1$ between ``Score'' and ``Quality'' (i.e. DockQ). Both ``Score'' and ``Quality'' occur within the range $0$ to $1$. These points were randomly shifted along the ``Score'' axis until $\rho = -0.1$. After randomly shifting the data, the scores were normalized to fall within the range $0 < {\rm Score} < 1$.  This synthetic dataset was balanced between positives and negatives using a positive cutoff of ${\rm Quality} \geq 0.23$ to mimic the DockQ cutoff for positive models. We then balanced the data set to have the same number of positive and negative models. 

We find that the fraction $\mu$ of the negative models with a ``Score'' less than the minimum Score of the positive models determines the large-sample limit for the hit rate plot. We start with $\mu = 0$ for the synthetic data set. To increase $\mu$, we apply Gaussian random shifting to the Scores of the negative models to increase their variance relative to the positive models. The Gaussian random shifts in the score are terminated when $\mu$ is within $10^{-7}$ of the target value. We find that when $
\mu > 0$, the hit rate plots in row (C) of Fig. \ref{fig:supp_synthetic_hr} begin to approach 0 as sample size $N_m$ increases. As $\mu$ increases, the hit rate plot diverges further as can be seen in the last column of row (c). 

\section{Physical features of protein interfaces}
\label{app:d}

%% Figure 12

\begin{figure*}[htbp]
\centering
\includegraphics[width=0.95\textwidth]{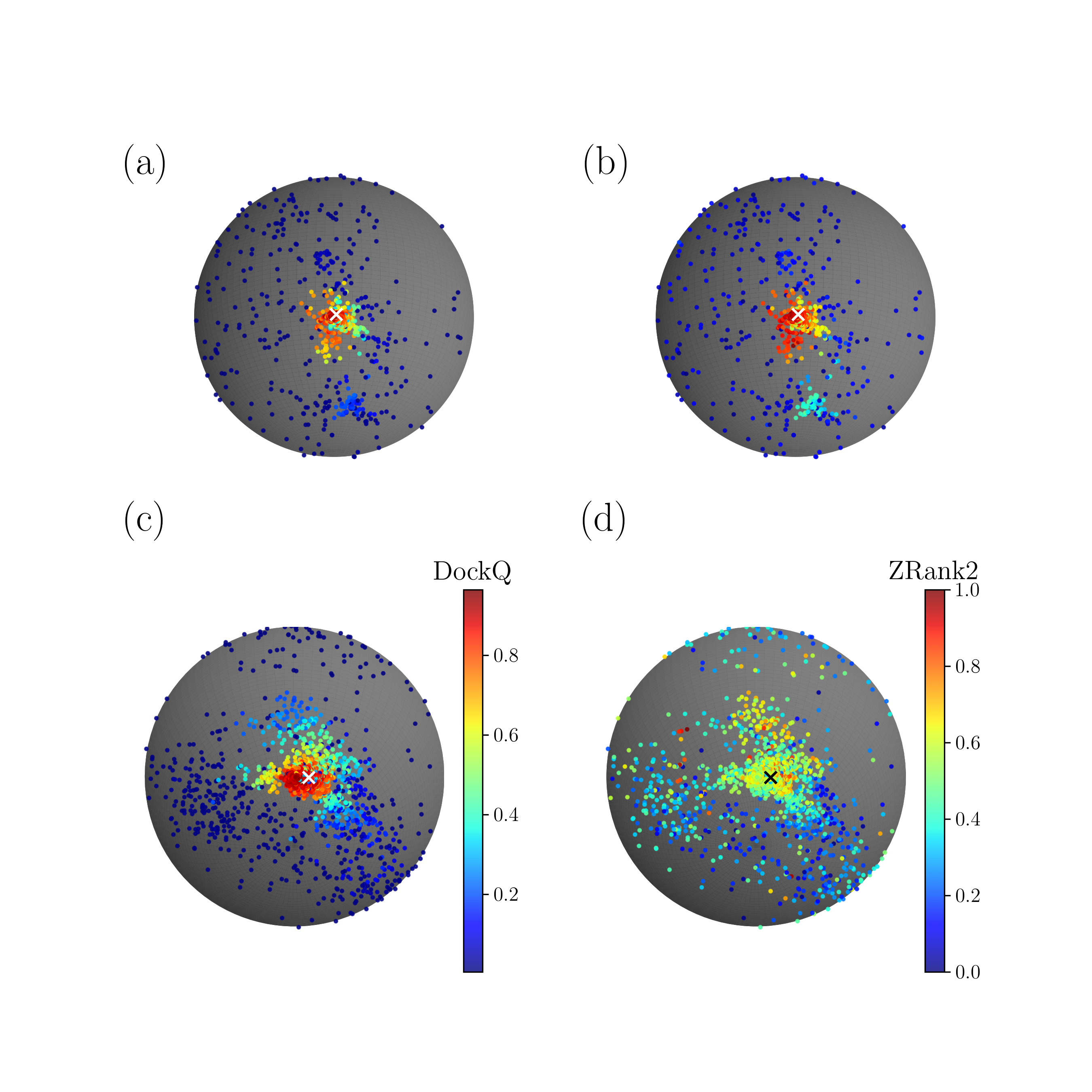}
\caption{The difficulty in scoring models can be determined by comparing their DockQ and PPI score landscapes. We show the center of mass positions (in spherical coordinates) of the C$_{\alpha}$ atoms of the ligands from models for 3WHQ shaded by (a) DockQ and (b) the rescaled ZRANK2 score with $0 \le {\rm ZRANK2} \le 1$. The score increases from dark blue to dark red. We show similar data for the ligand models for 2GRN shaded by (c) DockQ and (d) the rescaled ZRANK2 score. The crosses indicate the center of mass positions of the ligands in the x-ray crystal structures.}
\label{fig:app_easy_hard_sphere}
\end{figure*}

In this appendix, we examine the difficulty in scoring models for each target based on the physical features of the protein-protein interfcae.  We first compare the DockQ \cite{basu_dockq_2016} and PPI score, such as ZRank2 \cite{pierce_zrank_2007, pierce_combination_2008}, landscapes in Fig.~\ref{fig:app_easy_hard_sphere}, to investigate target scroing difficulty.  The landscapes are calculated as follows. The heterodimer model is split into its constituent monomers. One monomer is labeled as the ``receptor'' and the other is labeled as the ``ligand''. The heterodimer is translated so that the center of mass (COM) of the receptor is at the origin. The model is then rotated to align the receptor C$_\alpha$ atoms with those of the receptor in the x-ray crystal structure target. The points in the landscape represent the positions of each model ligand's COM in spherical coordinates with radial position $R$, given by the distance between the centers of mass of the receptor and the ligand in the x-ray crystal structure target.

In Fig.~\ref{fig:app_easy_hard_sphere}, we compare the DockQ and ZRank2 landscapes for ``easy'' and ``hard'' targets, 3WHQ and 2GRN, respectively. For 3WHQ and 2GRN, the DockQ landscapes are provided in panels (a) and (c), and the ZRank2 landscapes are provided in panels (b) and (d). The raw ZRank2 scores were negated and scaled so that $0 \le {\rm ZRANK2} \le 1$, where $1$ should give the most native-like model. For the ``easy'' target 3WHQ, the ZRank2 landscape is very similar to the DockQ landscape, whereas the ZRank2 landscape does not overlap with the DockQ landscape for the ``hard'' target 2GRN. 

%% Figure 13

\begin{figure*}[htbp]
\centering
\includegraphics[width=0.95\textwidth]{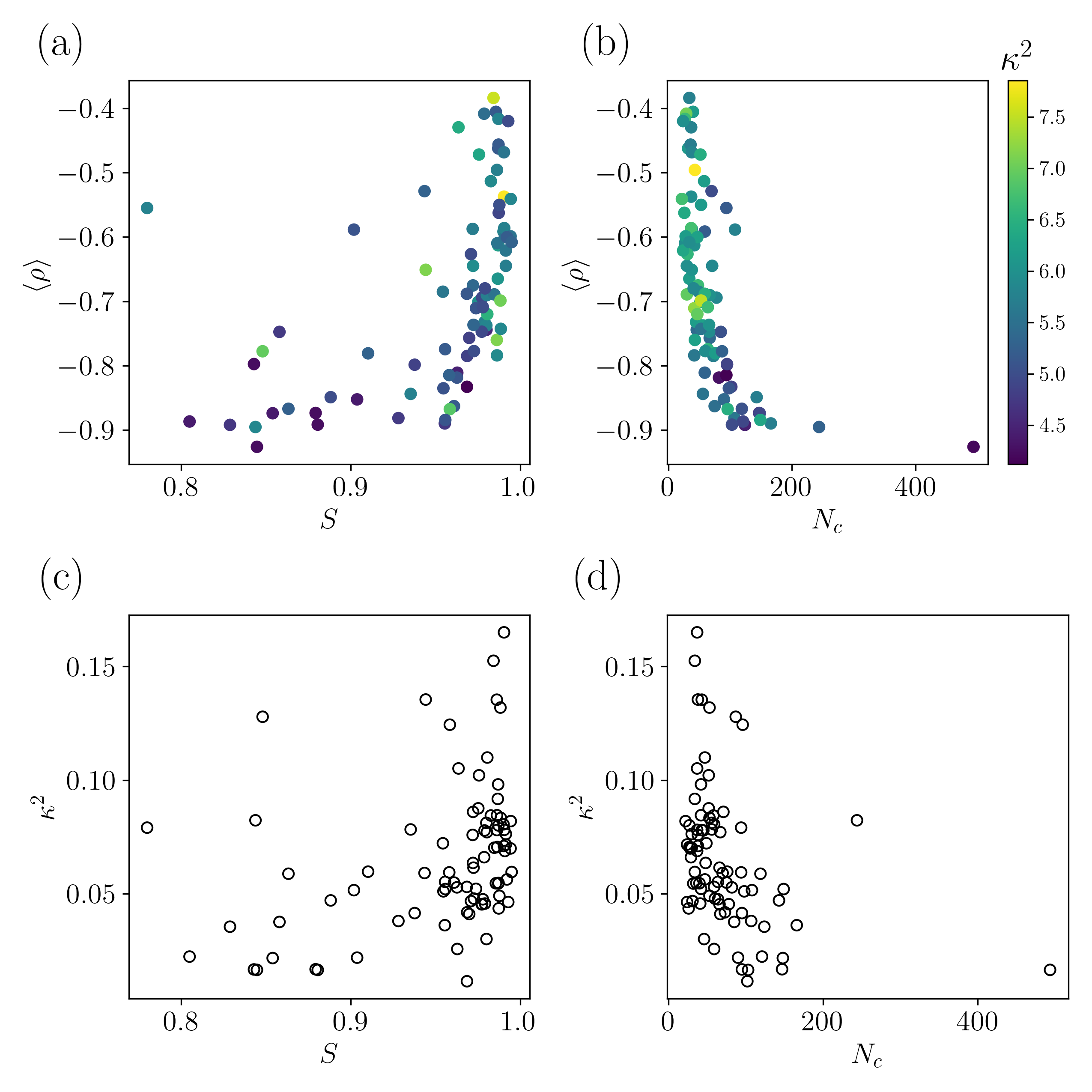}
\caption{Anisotropy of the DockQ landscapes correlates with physical features of the interfaces. The Spearman correlation coefficient $\langle \rho \rangle$ averaged over scoring functions for each target plotted versus (a) the interface separability $S$ and (b) the number of interfacial contacts $N_c$. The data points are shaded by the relative anisotropy $\kappa^2$ (in Eq.~\ref{eq:relative_anisotropy}) of the DockQ landscape increasing from dark blue to yellow.  $\kappa^2$ is also plotted versus (c) $S$ and (d) $N_c$ for each target.}
\label{fig:supp_anisotropy_vs_physical}
\end{figure*}

%% Figure 14

\begin{figure}[htbp]
\centering
\includegraphics[width=\columnwidth]{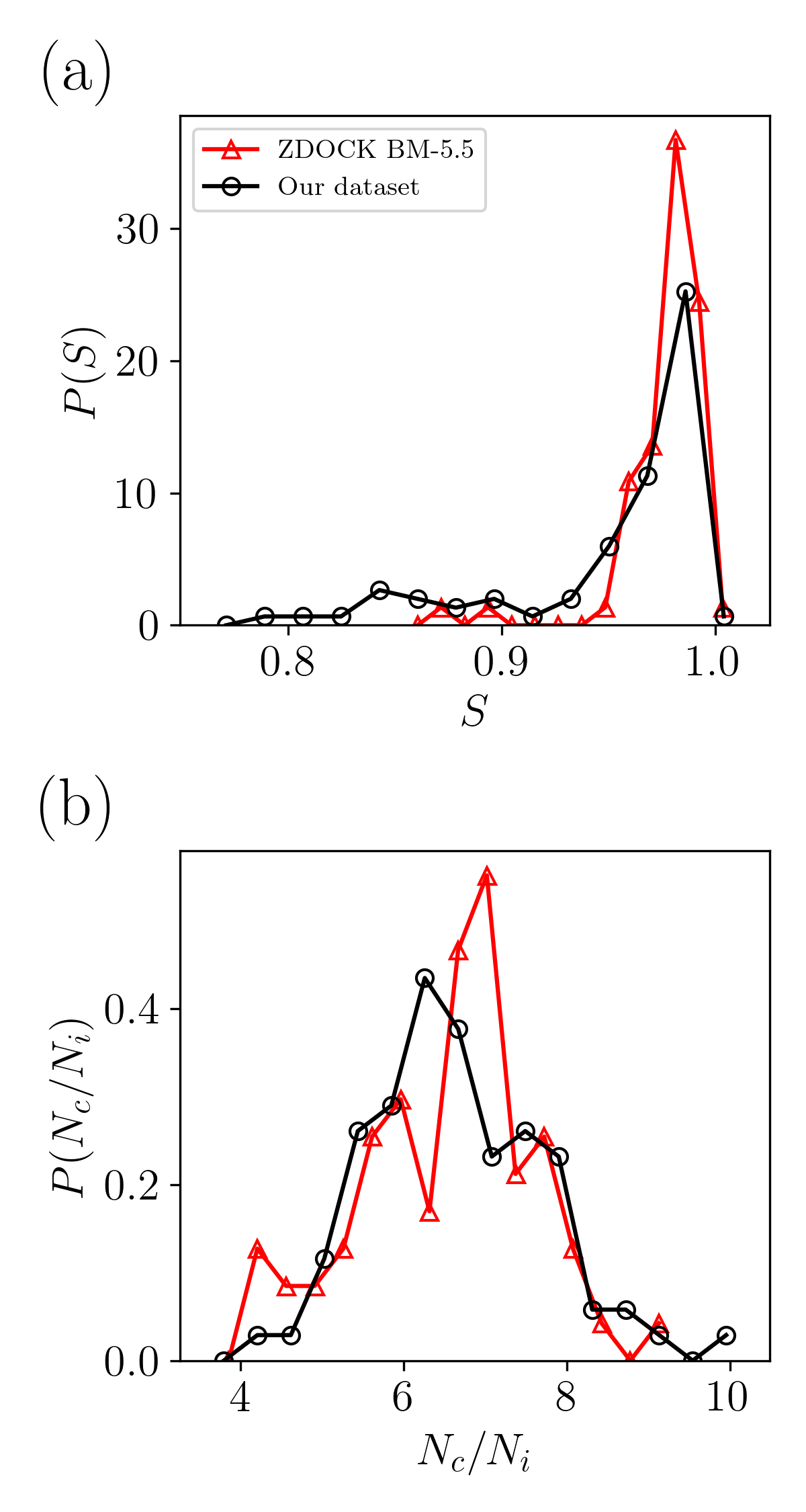}
\caption{The probability distributions of the physical features (a) $S$ and (b) $N_c / N_i$ for the $84$ targets in the main dataset (black circles) compared to the distributions for the ZDOCK Benchmark 5.5 dataset (red triangles), which contains $62$ heterodimers that do not overlap with the original dataset.}
\label{fig:bm55_comparison}
\end{figure}

Are attributes of the DockQ landscape correlated with physical features of protein interfaces? In Figure~\ref{fig:supp_anisotropy_vs_physical} (a) and (b), we compare $\langle \rho \rangle$, $S$, and $N_c$ to the relative anisotropy of the DockQ landscape for each target in the main $84$-target dataset. We show that the relative anisotropy $\kappa^2$ (in Eq.~\ref{eq:relative_anisotropy}) typically decreases as $|\langle \rho \rangle|$ increases and $S$ decreases or $N_c$ increases. We scatter $\kappa^2$ versus $S$ and versus $N_c$ in Fig.~\ref{fig:supp_anisotropy_vs_physical} (c) and (d), which indicate correlations between the shape of the DockQ landscape and physical features of protein interfaces.

We also compare the distributions of the physical features for the main dataset of $84$ heterodimer targets to the distributions of the physical features for the ZDOCK Benchmark 5.5 dataset with $62$ non-redundant heterodimer targets~\cite{vreven_updates_2015}. We plot the distributions of the physical features $S$ and $N_c/N_i$ in Fig.~\ref{fig:bm55_comparison} (a) and (b). We divide $N_c$ by the number of amino acids in the interface $N_i$ so that this feature does not depend on the size of the interface. We find similar distributions for the main $84$ taraget dataset (black circles) and the ZDOCK Benchmark 5.5 dataset (red triangles) for both physical features.

\clearpage
\bibliography{ppi_pre_references_cleaned}

\end{document}